\documentclass[aip,pof,preprint,amsmath,amssymb]{revtex4-1}
\usepackage{graphicx, subfigure}  
\usepackage{dcolumn}   
\usepackage{bm}        
\usepackage{amssymb}   
\usepackage{color} 
\date{\today}
\begin{document}
\title{Oscillatory instability and fluid patterns in low-Prandtl-number   Rayleigh-B\'{e}nard convection with uniform rotation}
\author{Hirdesh K. Pharasi} 
\affiliation{Department of Physics and Meteorology, Indian Institute of Technology, Kharagpur-721 302, India}  
\author{Krishna Kumar}
\affiliation{Department of Physics and Meteorology, Indian Institute of Technology, Kharagpur-721 302, India} 
\begin{abstract}
We present the results of direct numerical simulations of flow patterns in a low-Prandtl-number ($Pr = 0.1$) fluid above the onset of oscillatory convection in a Rayleigh-B\'{e}nard system rotating uniformly about a vertical axis. Simulations were carried out in a periodic box with thermally conducting and stress-free top and bottom surfaces. We considered a rectangular box ($L_x \times L_y \times 1$) and a wide range of Taylor numbers ($750 \le Ta \le 5000$) for the purpose. The horizontal aspect ratio $\eta = L_y/L_x$ of the box was varied from $0.5$ to $10$. The primary instability appeared in the form of two-dimensional standing waves for shorter boxes ($0.5 \le \eta < 1$ and $1 < \eta < 2$). The flow patterns observed in boxes with $\eta = 1$ and $\eta = 2$ were different from those with $\eta  < 1$ and $1 < \eta < 2$. We observed a competition between two sets of mutually perpendicular rolls at the primary instability in a square cell ($\eta = 1$) for $Ta < 2700$, but observed a set of parallel rolls in the form of standing waves for  $Ta \geq 2700$. The three-dimensional convection was quasiperiodic or chaotic  for $750 \le Ta < 2700$, and then bifurcated into a two-dimensional periodic flow for $Ta \ge 2700$. The convective structures consisted of the appearance and disappearance of straight rolls, rhombic patterns, and wavy rolls inclined at an angle $\phi = \frac{\pi}{2} - \arctan{(\eta^{-1})}$ with the straight rolls. 
\end{abstract}
\maketitle

\section{Introduction}
Rayleigh-B\'{e}nard system rotating uniformly about a vertical axis~\cite{rotating_rbc, veronis_jfm, chandrasekhar:book_1961, rossby_1969, dikarev_1983, knoblock_1998, clever_busse_2000, scheel_cross_pre_2005, prosperetti_2012} is an interesting problem for studying pattern-forming instabilities and bifurcations,~\cite{kuppers_lortz, kuppers, busse_heikes_1980, niemela_donnelly_1986, zhong_ecke_1991, rihai_prs_A_1992, patterns_rot_rbc, knoblock_silber_physicaD_1993, hu_etal, bajaj_etal_prl_1998, dawes_phy_lett_A_1999, dawes_physicaD_2000, guarino_vidal_pre_2004, sanchez_etal_pre_2005, scheel_etal, cox_mathieus_jfm_2000}  chaos,~\cite{cross_etal_pre_2001, rubio_etal_jfm_2010} and turbulence~\cite{turbulence_rot_rbc} in addition to its potential applications in rotating cryogenics.~\cite{cyogenics_applications} The rotation about a vertical axis introduces Coriolis force as well as centrifugal force~\cite{becker_etal_2006} in the flow. Rubio, Lopez \& Marques~\cite{rubio_etal_jfm_2010} showed interesting effects of the centrifugal force  even for small values ($\sim 10^{-2}$) of the Froude number $Fr$, which is a ratio of the centrifugal force to the force of buoyancy. The Coriolis force is known to break the mirror symmetry of the convective flow patterns~\cite{veronis_jfm, knoblock_1998} even at small rotation rates. Chandrasekhar~\cite{chandrasekhar:book_1961} analyzed the effects of Coriolis force on the onset of convection in a Rayleigh-B\'{e}nard system. The linear theory of Chandrasekhar predicted the onset of Rayleigh-B\'{e}nard convection (RBC) through a stationary instability for fluids with Prandtl number $Pr > 0.667$ for all rotation rates. The selection of patterns is, however, a purely nonlinear phenomenon, and cannot be determined  by a linear theory. The non-linear analysis of the problem by K\"{u}ppers and Lortz~\cite{kuppers_lortz} (KL) showed that (i) a set of parallel rolls was the steady state solution at the instability onset for the dimensionless rotation rates ($\Omega$) below a critical value $\Omega_{\star} (Pr)$, and (ii) the convection was always unsteady at the instability onset for $\Omega \ge \Omega_{\star}$. This led to interesting pattern dynamics. A set of parallel rolls was found to be replaced by another set of parallel rolls oriented at an angle of about $60^{\circ}$ with the old rolls~\cite{busse_heikes_1980, niemela_donnelly_1986, zhong_ecke_1991, cox_mathieus_jfm_2000} for $\Omega \ge \Omega_{\star}$.  The change in orientation of rolls was because of noise.  The linear theory of Chandrasekhar also predicted the possibility of oscillatory convection at the onset for $Pr < 0.677$  at dimensionless rotation rates above a critical value $\Omega_c (Pr)$. For a fluid with $Pr = 0.63$, the critical value $Ta_c$ of the Taylor number $Ta = 4 \Omega^2$ is  equal to $2.58 \times 10^5$. However, the oscillatory convection may appear at small or moderate rotation rates for low-Prandtl-number fluids. The previous investigations on instabilities and bifurcations above the onset of oscillatory Rayleigh-B\'{e}nard convection in fluids rotating uniformly about a vertical axis~\cite{rihai_prs_A_1992,  knoblock_silber_physicaD_1993, knoblock_1998, clever_busse_2000, dawes_phy_lett_A_1999, dawes_physicaD_2000} were done in small simulation boxes. 

Low-Prandtl-number convection~\cite{clever_busse_1981, chiffaudel_etal_1987,  thual_1992, gillet_etal_2007, pal_etal_epl_2009, low_P_stress_free} is relevant for geophysical~\cite{glatzmaier_etal} and astrophysical~\cite{cattaneo_etal} problems. The estimated value of $Pr$ for Earth's molten core~\cite{gillet_etal_2007} is approximately equal to $0.1$. The molten outer core and the inner solid core are separated by a transition zone, which itself is in liquid state. The other transition zone between the molten core and the lower mantle is also highly viscous and deformable. The transition zones on the two sides of Earth's molten core have discontinuities, and they keep drifting non-uniformly. The boundary conditions for the velocity and the temperature fields are complex on these bounding surfaces. The boundaries of the ``convective zone" in stars ($Pr \approx 10^{-8}$) are  also complex.  The numerical simulations to investigate the phenomenon of thermal convection in Earth's molten core, or in the convective zones of a star, are usually done either with {\it no-slip} or {\it free-slip} ({\it stress-free}) boundary conditions on the velocity field. The no-slip conditions are appropriate where the bounding surfaces are rigid and stationary. The predictions of the simulations with no-slip boundary conditions have the advantage of being verified in controlled laboratory experiments. However, the stress-free boundary conditions may be a more useful approximation on a boundary between two liquids with a large difference in their viscosities. Goldstein and Graham~\cite{goldstein_graham_1969} achieved  almost stress-free boundaries in experiments. The famous KL instability~\cite{kuppers_lortz} was predicted and understood for the first time by considering stress-free boundary conditions. In addition, the idealized boundary conditions are useful in developing simple models to investigate details of bifurcations qualitatively. Both types of boundary conditions are useful for better understanding of the natural convection in geophysical and astrophysical problems in the absence of exact boundary conditions. 

We present, in this article, the results of our investigations of the role of Coriolis force on the convective instabilities near the onset of  Rayleigh-B\'{e}nard convection (RBC) in a low-Prandtl-number fluid rotating uniformly about a vertical axis. We have considered $Pr = 0.1$, which is relevant for Earth's molten core. The onset of convection is then oscillatory~\cite{chandrasekhar:book_1961} at relatively smaller rotation rates. Our aim is to investigate the effects of the Coriolis force on the onset of RBC when the conduction state becomes unstable via an oscillatory instability. We have carried out direct numerical simulations (DNS) of three-dimensional flows in a  box with thermally conducting and stress-free top and bottom surfaces for the purpose. The stress-free boundary conditions also allowed us to investigate the effects of Coriolis force on the instabilities and bifurcations~\cite{thual_1992, low_P_stress_free} observed in low-Prandtl-number fluids without rotation. Simulations were done in a box of size $L_x \times L_y \times 1$ for a wide range of Taylor numbers ($750 \le Ta \le 5000$). The ratio $\eta = L_y/L_x$ of the horizontal dimensions $L_y$ and $L_x$ of the simulation box was varied in a range of $\eta$ ($0.5 \le \eta \le 10$).  This enabled us to investigate the nonlinear interactions between two sets of rolls of different wavelengths and at different orientations. The simulations done show the appearance of convection in the form of two-dimensional standing waves in smaller boxes ($0.5 \le \eta < 1$ and $1 < \eta < 2$). 

We observe fluid patterns, which vary quasiperiodically in time, at the instability onset in simulation boxes with $\eta = 1$ and $\eta = 2$. We observe three-dimensional (3D) patterns just above the onset of convection for $750 \le Ta \le 1114$ in a simulation box with square cross-section ($\eta = 1$). These patterns describe a nonlinear superposition of two sets of mutually perpendicular rolls. The two largest Fourier modes $W_{101}$ and $W_{011}$ show amplitude modulation with time. The amplitude and the frequency of modulation of both the modes are found to be equal. In a small window of Taylor numbers ($1115 \le Ta < 1125$), the primary instability displays temporally chaotic behavior. The chaotic flow becomes once again quasiperiodic at $Ta = 1125$. However, the amplitude and the frequency of modulation of both the modes $W_{101}$ and $W_{011}$ are unequal in this case. The resulting patterns are cross-rolls varying quasiperiodically in time. The amplitude of $W_{011}$ decreases with further increase in $Ta$. A bifurcation from three-dimensional quasiperiodic flow to two-dimensional (2D) periodic convection occurs at the primary instability for $Ta = 2700$. The 2D oscillatory rolls continue to exist until $Ta = 5000$. We also present a low-dimensional model constructed for a square simulation box. In a box with rectangular cross-section ($\eta = 2$), we observe the appearance of convection as 3D waves varying quasiperiodically in time for all values of $Ta$ considered here. Three-dimensional (3D) temporally chaotic convection is observed at the instability onset in longer boxes ($2 < \eta \le 10$). The convective patterns at the onset show a chaotic competition between two sets of rolls oriented at an angle $\phi = \frac{\pi}{2} - \arctan{(\eta^{-1})}$, as observed in KL instability. However, the sequence of convective patterns during KL type instability consist of rolls, rhombic patterns, and oblique wavy rolls just above the primary instability. 

\section{Hydrodynamical System}
We consider a thin layer of a Boussinesq fluid of mean density $\rho_{\circ}$, thermal expansion coefficient $\alpha$, kinematic viscosity $\nu$ and thermal diffusivity $\kappa$ confined between two conducting horizontal surfaces separated by a distance $d$. The whole system is supposed to be rotating uniformly about a vertical axis with uniform rotation rate $\Omega_0$. The bottom surface is uniformly heated, while the top surface is uniformly cooled. An adverse temperature gradient $\beta =\Delta T/d$ is imposed across the fluid layer by maintaining a temperature difference $\Delta T$ across it. We have considered convection near the onset in a low-Prandtl-number ($Pr =0.1$) fluid with the Boussinesq parameter $\delta = \alpha \beta d \sim 10^{-2}$ for dimensionless rotation rate $27 < \Omega < 36$. The ratio of Coriolis force and the force of buoyancy is $\sqrt{Ta}/Ra \sim 10^{-2}$, while the ratio  of centrifugal force and Coriolis force is $\Omega Pr \delta \Gamma/ 2 \sim 10^{-2}$ for all values of $Ta$ considered here. All the simulations have been done in boxes of aspect ratio $\Gamma = L_y/d$ between $2.6$ and $11.2$. The Froude number $Fr = \delta Pr Ta \Gamma/ 4 Ra$ therefore varies from $3.5 \times 10^{-4}$ to $3.5 \times 10^{-3}$. The effects of centrifugal force are therefore ignored here. We have considered the state of steady conduction as the basic state in a rotating frame of reference. We have not considered the situation where the basic state has a mean flow. The dimensionless hydrodynamic equations, which govern the convective flow in a Boussinesq fluid, are then given by,
\begin{eqnarray}
\partial_t {\bf v} + ({\bf v \cdot \nabla}){\bf v} &=& -\nabla p + \nabla^2 {\bf v} + Ra\theta{\bf\hat{z}} +\sqrt{Ta}({\bf{v\times}{\bf{\hat{z}}}}),\label{eq:momentum}\\
 Pr[\partial_t {\theta} +({\bf v}{\cdot}\nabla)\theta] &=& {\nabla}^2 \theta + v_3,\label{eq:theta}\\
\nabla{\cdot}{\bf v} &=& 0, \label{eq:continuity}
\end{eqnarray}
\noindent
where ${\bf{v}}(x,y,z,t)=(v_{1},v_{2},v_{3})$ is the velocity field, and the unit vector ${\bf\hat{z}}$ is directed vertically upward. The deviations in the pressure and temperature fields from their values in the basic state of stationary conduction are denoted as $p(x,y,z,t)$ and $\theta(x,y,z,t)$ respectively. Lengths, time, velocity and temperature fields are measured in units of the fluid thickness $d$, the viscous diffusion time $d^2/\nu$, $\nu/d $ and  $\beta d \nu/\kappa$ respectively. Dimensionless numbers are:  Prandtl number $Pr=\nu/\kappa$, Rayleigh number $Ra = g\alpha \beta d^{4}/{\nu \kappa}$ and Taylor number $Ta$ $=$ $4\Omega^{2}_{0} d^{4}/\nu^{2}$ $=$ $4\Omega^{2}$, where $g$ is the gravitational acceleration. The boundary conditions at thermally conducting and stress-free bounding surfaces located at $z=0,~1$ are:
\begin{equation}
\partial_{z}v_{1} = \partial_{z}v_{2} = v_{3}=\theta=0 \text{ at } z=0,1.\label{bcs}
\end{equation}
\noindent
All fields are considered to be periodic in horizontal plane. The use of periodic boundary conditions is very popular to simulate flows in an extended layer of fluid. It captures the flow structure nicely away from the lateral walls. The periodic boundary conditions (PBC) allow a high resolution simulation in a box. The simulation of a flow in a large container without PBC may require prohibitive grids to resolve the flow structure. The use of PBC is the most appropriate with pseudo spectral method used here, as it facilitates the computation of terms like ${\nabla}^2 \theta$, ${\nabla}^2 {\boldmath{v}}$ exactly in numerical sense.  

The nature of thermal convection at the primary instability~\cite{chandrasekhar:book_1961} depends upon the values of the Taylor number $Ta$ and Prandtl number $Pr$. The critical Rayleigh number $Ra_{c}(Ta)$ and the corresponding wavenumber $k_{c}(Ta)$ for stationary convection are:
\begin{eqnarray}
Ra_{c}(Ta) &=& \frac{1}{k^2_{c}}\left[ (\pi^2+k^2_{c})^3+\pi^2 Ta\right],~~\   
k_{c}(Ta) = \pi \sqrt{a_{+}+a_{-}-1/2},\nonumber\\ 
\text{with }  a_{\pm} &=& \left( \frac{1}{4}\left\lbrace \frac{1}{2}+\frac{Ta}{\pi^4}\pm \left[ \left( \frac{1}{2} +\frac{Ta}{\pi^4}\right)^2 -\frac{1}{4}\right]^\frac{1}{2} \right\rbrace \right)^\frac{1}{3}.
\end{eqnarray}
The critical Rayleigh number $R_{\circ} (Ta, Pr)$ and the corresponding wavenumber $k_{\circ} (Ta, Pr)$ for the oscillatory convection are: 
\begin{eqnarray}
Ra_{\circ} (Ta, Pr) &=& 2\left( \frac{1+Pr}{k^2_{\circ}}\right) \left[ (\pi^2+k^2_o)^3+ \frac{\pi^2 Ta Pr^2}{(1+Pr)^2}\right],~~\  
k_{\circ} (Ta, Pr) = \pi \sqrt{b_{+}+b_{-}-1/2},\nonumber\\
\text{with } b_{\pm} &=& \left( \frac{1}{4}\left\lbrace \frac{1}{2}+\frac{Pr^2}{(1+Pr)^2}\frac{Ta}{\pi^4}\pm  \left[ \left( \frac{1}{2}+\frac{Pr^2}{(1+Pr)^2}\frac{Ta}{\pi^4}\right)^2 -\frac{1}{4}\right]^\frac{1}{2} \right\rbrace \right)^\frac{1}{3}.
\end{eqnarray}
The angular frequency $\omega_{\circ} (Ta, Pr)$ of the oscillatory convection is given by the expression:
\begin{equation}
\omega_{\circ} (Ta, Pr) = \left[ \frac{\pi^2 Ta}{(\pi^2+k^2_{\circ})} \left( \frac{1 - Pr}{1 + Pr} \right)-\left( \pi^2 + k^2_{\circ} \right)^2 \right]^{\frac{1}{2}}.
\end{equation} 
It is always real, and $Ra_{\circ} (Ta, Pr) < Ra_c (Ta)$ in a fluid with $Pr = 0.1$, if $Ta > Ta_{*} = 728$. The conduction state therefore loses stability via oscillatory instability, if $Ta > 728$ in a fluid with $Pr = 0.1$.

\section{Numerical simulations}
The velocity, temperature, and pressure fields are expanded as:
\begin{eqnarray}
v_1 (x,y,z,t) &=& \sum_{l,m,n} U_{lmn}(t) e^{i(lk_xx+mk_yy)} \cos{(n\pi z)},\label{eq.u}\\
v_2 (x,y,z,t) &=& \sum_{l,m,n} V_{lmn}(t) e^{i(lk_xx+mk_yy)} \cos{(n\pi z)},\label{eq.v}\\
v_3 (x,y,z,t) &=& \sum_{l,m,n} W_{lmn}(t) e^{i(lk_xx+mk_yy)} \sin{(n\pi z)},\label{eq.w}\\
\theta(x,y,z,t) &=& \sum_{l,m,n} \Theta_{lmn}(t) e^{i(lk_xx+mk_yy)} \sin{(n\pi z)},\label{eq.th}\\
p (x,y,z,t) &=& \sum_{l,m,n} P_{lmn}(t) e^{i(lk_xx+mk_yy)} \cos{(n\pi z)},\label{eq.p}
\end{eqnarray} 
where $U_{lmn}(t)$, $V_{lmn}(t)$, $W_{lmn}(t)$,  $\Theta_{lmn}(t)$, and $P_{lmn}(t)$ are the Fourier amplitudes in the expansion of the fields $v_{1}$, $v_{2}$, $v_{3}$, $\theta$, and $p$ respectively. ${\bf k} = k_x \hat{x} + k_y \hat{y}$ is the horizontal wave vector of the perturbations.  The integers $l,m,n$ can take values consistent with the continuity equation~(Eq.~\ref{eq:continuity}). This leads to the condition:
\begin{equation} 
lk_x U_{lmn} + mk_y V_{lmn} + n \pi W_{lmn} =0. \label{cont_1}.
\end{equation}
We allow all the modes, which are compatible with Eq.~\ref{cont_1}, in our simulation. Perturbations with the critical wavenumber $k_{\circ} (Ta, Pr)$ for the oscillatory convection are the most dangerous ones, as soon as $Ra$ is raised above its critical value $Ra_{\circ} (Ta, Pr)$. In addition, perturbations with long wavelengths are always present in larger containers. We therefore set $k_x = k_{\circ} $, and treat $k_y = q$ as a parameter. Simulations are done in a rectangular box of size $L_x \times L_y \times 1$, where $L_x = 2\pi/k_{\circ}$ and $L_y = 2\pi/q$. The ratio  $ \eta = L_y/L_x = k_{\circ}/q$ is varied in a wide range ($0.5 \le \eta \le 10$) for each value of $Ta$. Varying the parameter $\eta$ allows us to investigate the nonlinear interactions of perturbations of different wavelengths with 2D rolls.  We numerically integrate the hydrodynamic system (Eqs.~\ref{eq:momentum}-\ref{eq:continuity}) with the boundary conditions (Eq.~\ref{bcs}) using pseudo-spectral method.~\cite{tarang} The code is validated for low-Prandtl-number RBC with and without rotation. The limit of zero-Prandtl-number RBC~\cite{thual_1992} without rotation was investigated in detail by Pal et al.~\cite{pal_etal_epl_2009} after validating the code with the results of Thual~\cite{thual_1992} for stress-free boundary conditions. The turbulent flow in low-Prandtl-number RBC with uniform rotation was studied using the code by Pharasi et al.~\cite{pharasi_etal_pre_2011} after reproducing the results of King et al.~\cite{king_etal_nature_2009} for $Pr =1$. We have used three spatial grids ($64^3$, $64 \times 128 \times 64$ and $128^3$) for our simulations. A standard fourth order Runge-Kutta (RK4) scheme with step size  $\delta t = 0.0005$ was used for time advancement.  Simulations for different values of $Ta$ were carried out with small random initial conditions and were continued until a steady state was reached.  The Fourier modes $U_{00n}$ and $V_{00n}$, which are  responsible for the mean horizontal flow, are not generated in the simulations for all the values of $Ta$ considered here. The convective flow was assumed to be steady, if the temporal behavior of the box averaged kinetic energy $E = \frac{1}{2}<v_1^2+v_2^2+v_3^2>$ and the Nusselt number $Nu = 1 + Pr^2 <v_3 \theta>$ showed steady behavior for a period  more than three hundred viscous diffusion time $d^2/\nu$. The power spectra of the  Fourier modes $W_{101}$ and $W_{011}$ also confirmed the flow in a steady state. The symbol $< >$ stands for the average over the simulation box. The final values of all fields in a steady state, in a given run, were also used as initial conditions for the next run when the reduced Rayleigh number $r = Ra/ {Ra_{\circ}(Ta, Pr)}$ was raised in small steps ($0.005 \le \Delta r \le 0.1$) for a fixed value of $Ta$. The simulations showing temporally quasiperiodic states were first done on $64^3$ grids for more than  three hundred viscous diffusion time. Then the final values of all the fields were used as initial conditions on $128^3$ grids for more than one hundred dimensionless time unit. Similar spatial grids were used for longer simulation boxes.

\begin{table}
\caption{Convective patterns for $Pr =0.1$ just above the primary instability ($r = \frac{Ra}{Ra_{\circ} (Ta, Pr)} = 1.005$) in different simulation boxes. Two-dimensional (2D) and three-dimensional (3D) standing waves (SW),  quasiperiodic cross-rolls (QPCR), and K\"{u}ppers-Lortz (KL) type patterns were observed in DNS.}
\begin{ruledtabular}
\begin{tabular}{ccccc}
$\eta = L_{y}/L_{x}$ & $Ta =750$ & $Ta =1100$ &  $Ta =1130$ & $Ta =2700$ \\ \hline
1/2&2D~SW&2D~SW&2D~SW&2D~SW \\
4/7&2D~SW&2D~SW&2D~SW&2D~SW \\
4/5&2D~SW&2D~SW&2D~SW&2D~SW \\
1&QPCR&QPCR&QPCR&2D~SW \\
6/5&2D~SW&2D~SW&2D~SW&2D~SW \\
3/2&2D~SW&2D~SW&2D~SW&2D~SW \\
4/3&2D~SW&2D~SW&2D~SW&2D~SW \\
2&QPCR&QPCR&QPCR&QPCR \\
4&KL-type&KL-type&KL-type&KL-type \\
10&KL-type&KL-type&KL-type&KL-type \label{table1}\\
\end{tabular}
\end{ruledtabular}
\end{table}

\section{Results and discussions}
The convective patterns, computed from DNS, just $0.5 \%$ above the onset of convection are listed in Table~\ref{table1}. For shorter periodic boxes ($ 0.5 \le \eta  < 1$ and $1 < \eta < 2$), we observe the onset of convection in the form of two-dimensional (2D) standing waves (SW). The convective patterns consist of a set of straight rolls. The locations of up-flow and down-flow alternate periodically. The patterns computed by DNS in small simulation boxes are in agreement with earlier predictions~\cite{knoblock_silber_physicaD_1993, dawes_physicaD_2000} near the onset of oscillatory RBC with rotation. Dawes~\cite{dawes_physicaD_2000} called the time-periodic 2D rolls as standing rolls. The boxes with smaller cross-section in horizontal plane do not allow two sets of rolls simultaneously. However, we observe interesting behavior in square ($\eta=1$) and rectangular ($\eta =2$) boxes. The convection appears in the form of a quasiperiodic competition between two sets of mutually perpendicular sets of rolls. We label them as quasiperiodic cross-rolls (QPCR). Similar patterns are also known as quasiperiodic standing waves~\cite{knoblock_silber_physicaD_1993} or quasiperiodic standing cross-rolls~\cite{dawes_physicaD_2000} in literature. We also observe 
a chaotic competition between two sets of wavy rolls oriented with each other at an angle near the primary instability in larger simulation boxes ($4 \le \eta \le 10$) for all values of $Ta$ considered here. Rhombic patterns, which  are due to the nonlinear superposition of the two sets of wavy rolls, are observed during the change in orientation of the rolls.  We label them as KL-type patterns.    
 
\begin{figure}[h]
\begin{center}
\includegraphics[height=11 cm,width= 11.5 cm]{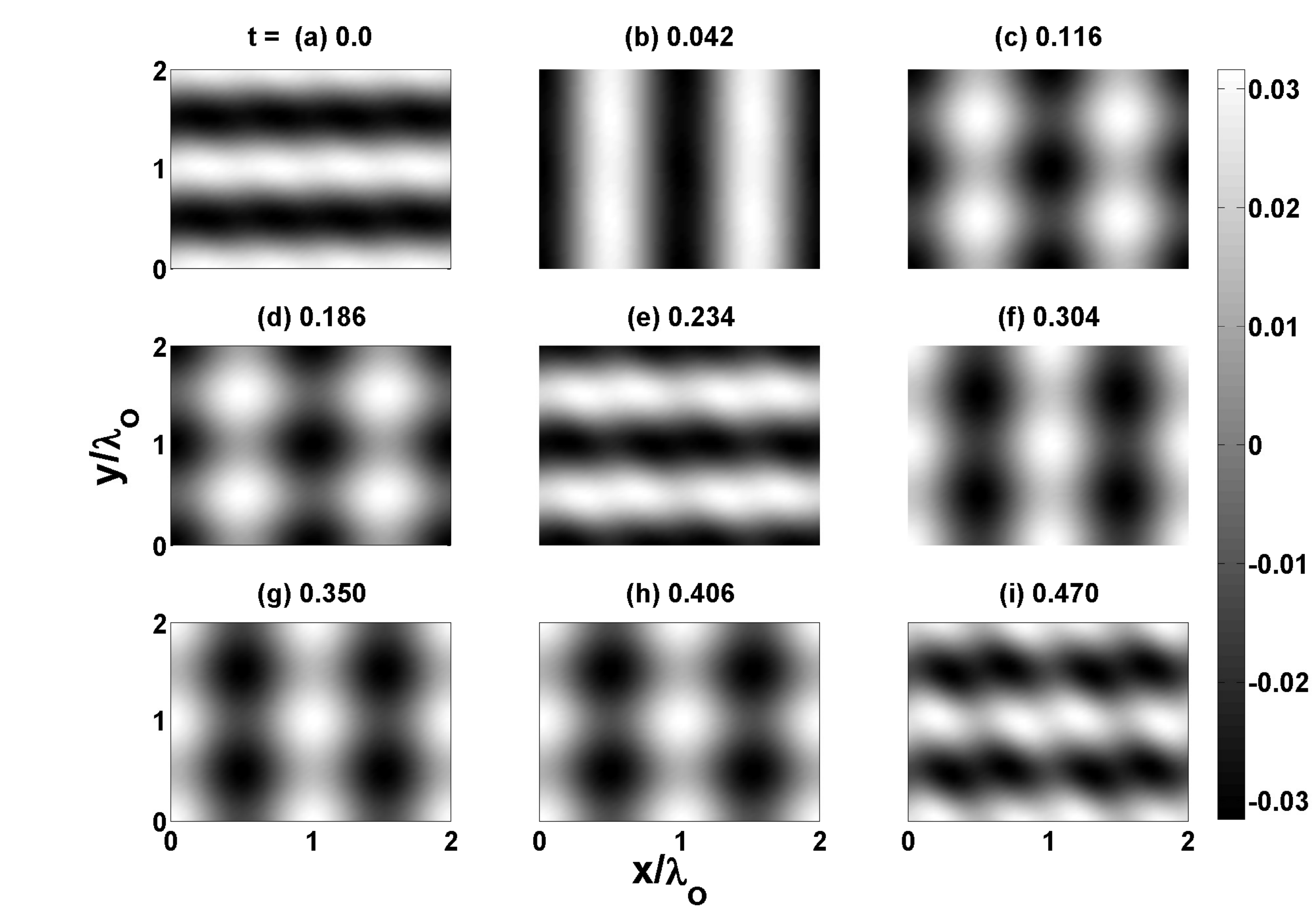}
\caption{The evolution of convective patterns with time in a square box ($\eta = 1$)-- Contour plots of the temperature field computed from direct numerical simulation (DNS) at the mid-plane ($z=0.5$) for $Ta = 750$, $Pr = 0.1$ and $r = Ra/Ra_{o} = 1.005$ at different instants of dimensionless time  t [(a)\---(i)]. It shows different phases of a time dependent competition of mutually perpendicular sets of rolls.} \label{contour_Ta_750} 
\end{center}
\end{figure}

\begin{figure}[h]
\begin{center}
\includegraphics[height=11 cm,width=14 cm]{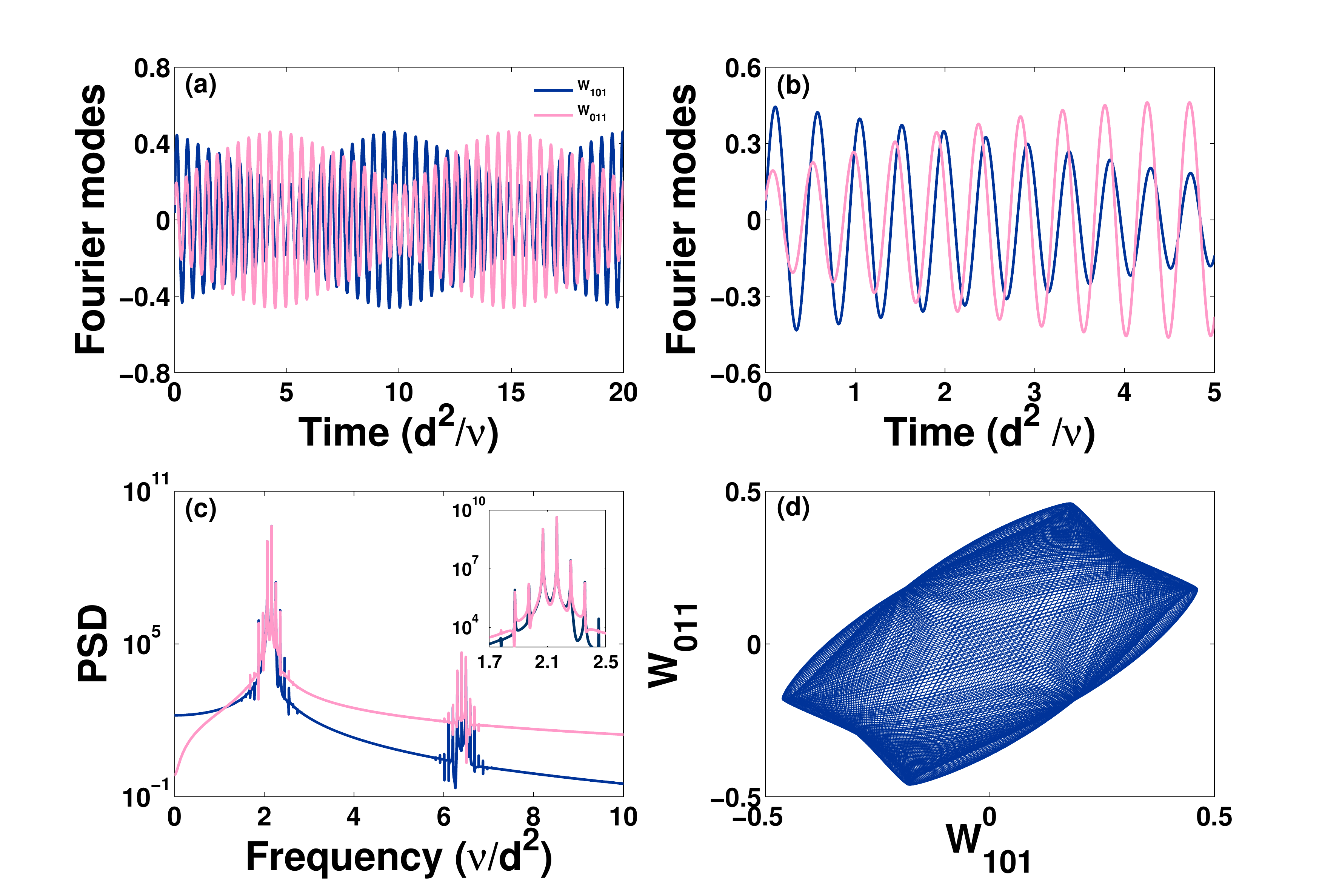}
\caption{(Color online) Properties of the two largest Fourier modes $W_{101}$ and $W_{011}$ for the patterns ($r = 1.005$, $Ta =750$, $\eta = 1$) shown in Fig.~\ref{contour_Ta_750}: (a) the variation of the Fourier modes  $W_{101}$ [blue (black) curves] and $W_{011}$ [pink (gray) curves] with time, (b) the same as in (a) but for a shorter time interval, (c) the power spectral density (PSD) of the modes $W_{101}$ [blue (black) curves] and $W_{011}$ [pink (gray) curves], and (d) the projection of the phase space on the $W_{101} - W_{011}$ plane.} \label{qp_Ta_750}
\end{center}
\end{figure}

\subsection{Patterns in a square box ($\eta = L_y/L_x = 1$)}
We now present the results computed from DNS in a box with square ($\eta = 1$) cross-section, and discuss them. The contour plots of the temperature field for $Pr = 0.1$ at different instants of time are displayed in Fig.~\ref{contour_Ta_750}. It shows different patterns due to time dependent competition between two sets of mutually perpendicular rolls. We observe square patterns whenever the amplitudes of two sets of rolls become equal. The patterns appear as cross rolls for most of the time. The positions of up-flow and down-flow alternate in time. 

The temporal variations of the two largest Fourier modes $W_{101}$ and $W_{011}$ for the patterns shown in Fig.~\ref{contour_Ta_750} are displayed in the first row of Fig.~\ref{qp_Ta_750} ($Ta = 750$ and $r = 1.005$). The  amplitudes and frequencies of modulation of both the modes $W_{101}$ and $W_{011}$ are equal, but they are out of phase [Fig.~\ref{qp_Ta_750}(a)]. The faster time variation of these two modes are compared in Fig.~\ref{qp_Ta_750} (b). The phase difference between the two modes  varies slowly with time. The power spectra [Fig.~\ref{qp_Ta_750} (c)] of the Fourier modes $W_{101}$ [blue (black) curve] and $W_{011}$ [pink (gray) curve] look similar. The peaks in the power spectra appear at exactly the same frequencies, although their heights are different. The power spectra [see the inset of Fig.~\ref{qp_Ta_750} (c)] show the two largest peaks at frequencies $f^{(1)}$ and $f^{(2)}$ ($ < f^{(1)}$), which differ slightly. The difference between the two frequencies $\Delta f = (f^{(1)} - f^{(2)})$ is  the frequency of amplitude modulation. There are many smaller peaks in the spectra at frequencies $f^{(1)} \pm j \Delta f$ and $f^{(2)} \pm j \Delta f$, where $j$ is a positive integer. Other frequencies are the higher harmonics of these frequencies. The projection of the phase space on the $W_{101} - W_{011}$ plane is shown in Fig.~\ref{qp_Ta_750} (d) after a long time. The region of the phase space gets continuously filled as time passes.  The trajectories in the actual ($128^3$ dimensional) phase space never return to any point. The behavior observed in Fig.~\ref{qp_Ta_750} (b) -(d) clearly shows that the convective patterns vary quasiperiodically in time. We label the patterns as quasiperiodically oscillating cross-rolls (QPCR). These patterns are consistent with the results of Rihai,~\cite{rihai_prs_A_1992} who ruled out any time-periodic three dimensional convection in square box. Similar patterns were also predicted by Dawes~\cite{dawes_physicaD_2000} in a low-Prandtl-number fluid ($ 0.15 < Pr < 0.2$) for much higher values of the Taylor number ($Ta > 3 \times 10^5$). We observe these patterns at the onset of convection for $0.07 < Pr < 0.2$ at much smaller rotation rates.  Similar patterns are also observed in numerical simulations~\cite{sanchez_etal_pre_2005} with no-slip boundary conditions.

\begin{figure}[h]
\begin{center}
\includegraphics[height=10 cm,width=16 cm]{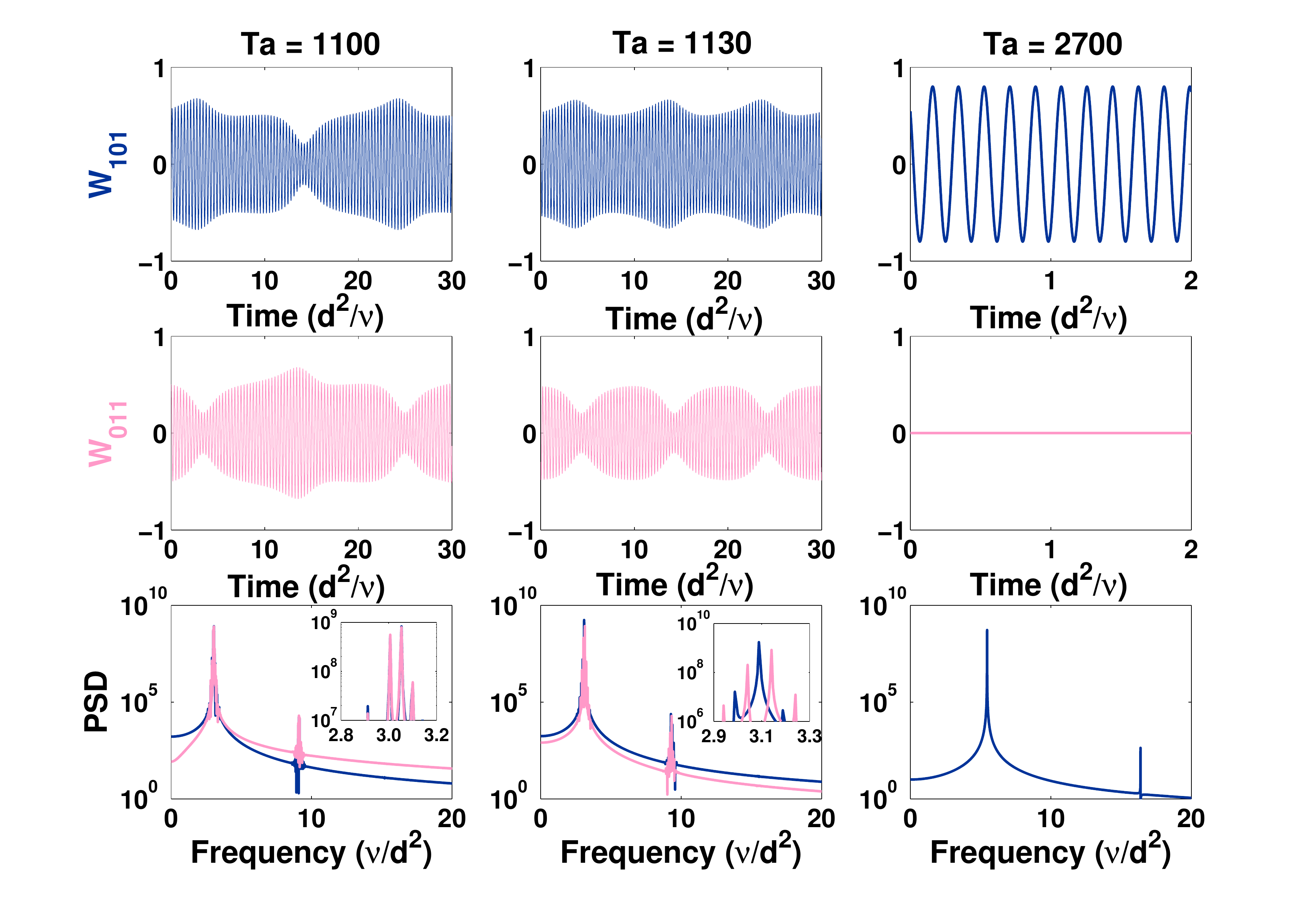}
\caption{(Color online) Bifurcations at the instability onset ($r = 1.005$) in a square box ($\eta =1$) with the variation of Taylor number $Ta$. The plots in the first, second and third columns are for $Ta = 1100$, $Ta = 1130$ and $Ta = 2700$ respectively. The temporal variation of the Fourier modes $W_{101}$ [blue (black) curves] is given in the first row for three different values of $Ta$, while the same is given for the Fourier mode $W_{011}$ [pink (gray) curves] in the second row. The third row shows the power spectral density for the modes $W_{101}$ [blue (black) curves] and $W_{011}$ [pink (gray) curves]
for three different values of $Ta$.} \label{psd_sq_box}
\end{center}
\end{figure}

As $Ta$ is raised, the temporal variation of these modes becomes more complex, but the convection remains quasiperiodic at the instability onset. The amplitude and the frequency of modulation of the Fourier modes $W_{101}$ and $W_{011}$ remain similar and out of phase for $750 \leq Ta \le 1114$. The period of amplitude modulation increases with $Ta$ for $750 \leq Ta \le 1114$.   
The first column of Fig.~\ref{psd_sq_box} shows properties of the two largest Fourier modes just above the onset ($r = 1.005$) for $Ta = 1100$. 
The Fourier modes become chaotic (not shown here) in a small window of $Ta$ ($1115 \le Ta < 1125$). These modes become temporally quasiperiodic once again for $1125 \leq Ta < 2700$. The first two rows of the second column of Fig.~\ref{psd_sq_box} display the temporal variations of the two largest modes for $Ta = 1130$ and $r = 1.005$. The amplitudes of modulation of the modes $W_{101}$ and $W_{011}$ become different. The largest  and the second largest peaks in the power spectra of these two modes appear at slightly different frequencies. However, the modulation frequency for both the modes remains the same. The third column of Fig.~\ref{psd_sq_box} describes the convection near onset for $Ta = 2700$. The Fourier mode $W_{011}$ vanishes and the power spectra of the Fourier mode $W_{101}$ shows the peak at a single frequency. This signifies a bifurcation from a three-dimensional (3D) quasiperiodic convection to a two-dimensional (2D) periodic convection. The state of 2D standing waves at the primary instability continues to exist until $Ta = 5000$.

\begin{figure}[h]
\begin{center}
\includegraphics[height=9 cm,width=16 cm]{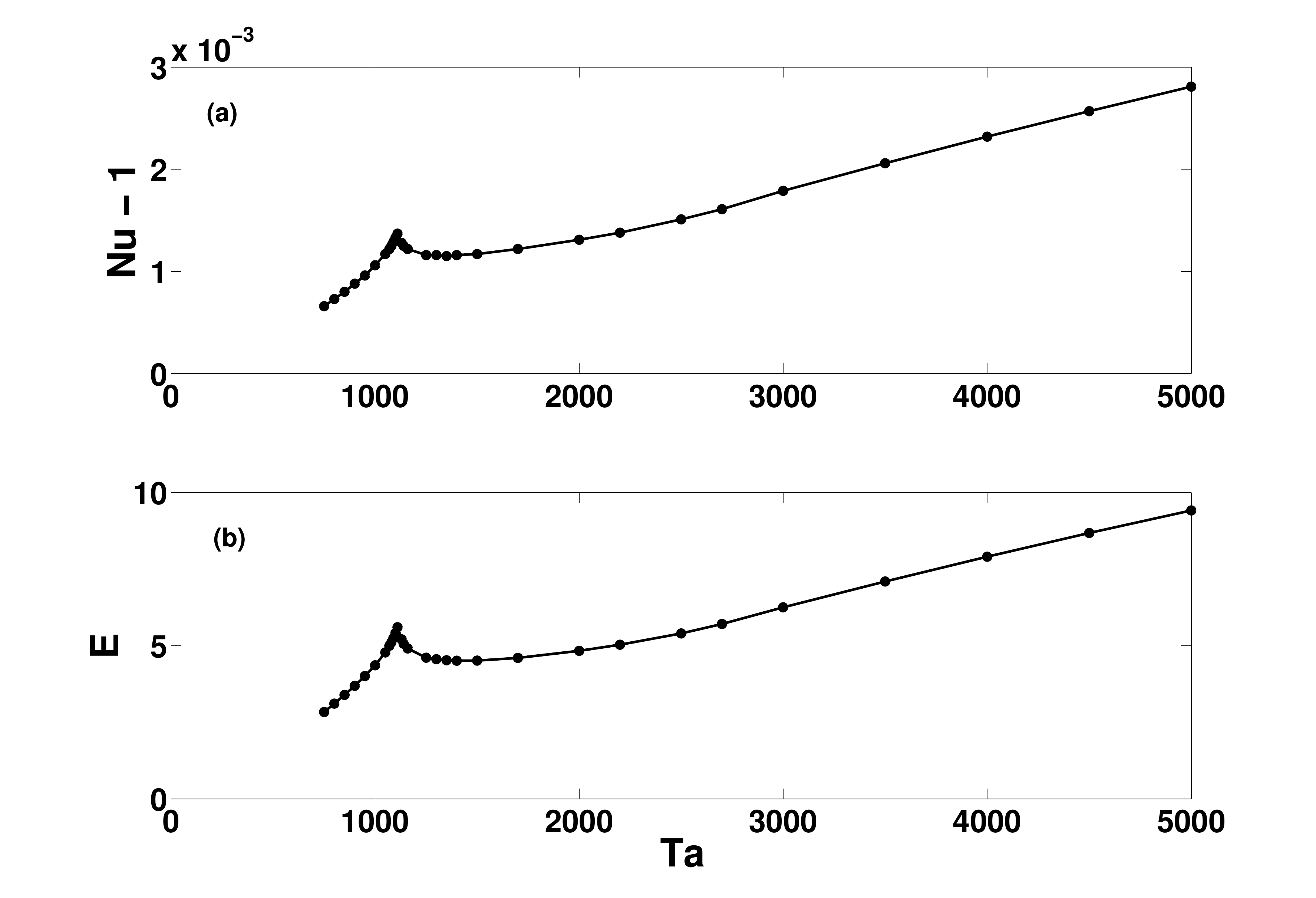}
\caption{The effect of rotation on the convective heat flux and the kinetic energy just above the primary instability ($r = 1.005$) for $Pr =0.1$ and $\eta = 1$: The variation of the maximum of (a) the Nusselt number $Nu -1$  and (b) the box averaged kinetic energy $E$ with the Taylor number $Ta$. Both $E$ and $Nu$ show a bifurcation at $Ta = 1114 \pm 0.5$.} \label{figure4}
\end{center}
\end{figure}

The variation of the maximum of convective heat flux ($Nu -1$) and box averaged kinetic energy $E$ with $Ta$ near the instability onset ($r = 1.005$) in a square box ($\eta = 1$) are shown for $Pr = 0.1$ in Fig.~\ref{figure4}. The heat flux due to convection $Nu-1$ remains small, as expected at smaller values of the Prandtl number $Pr$ [see, Fig.~\ref{figure4} (a)]. It increases initially with $Ta$, and reaches a peak at $Ta = 1114$. It then decreases slightly with increasing $Ta$, and again starts increasing monotonically with $Ta$. The  variation of the maximum of kinetic energy $E$ with $Ta$ [Fig.~\ref{figure4} (b)] shows exactly the same behavior. The  peaks at $Ta = 1114$ in the plots of both $Nu-1$ and $E$ with $Ta$ indicate a bifurcation.

\begin{figure}[h]
\begin{center}
\includegraphics[height=10 cm,width= 16 cm]{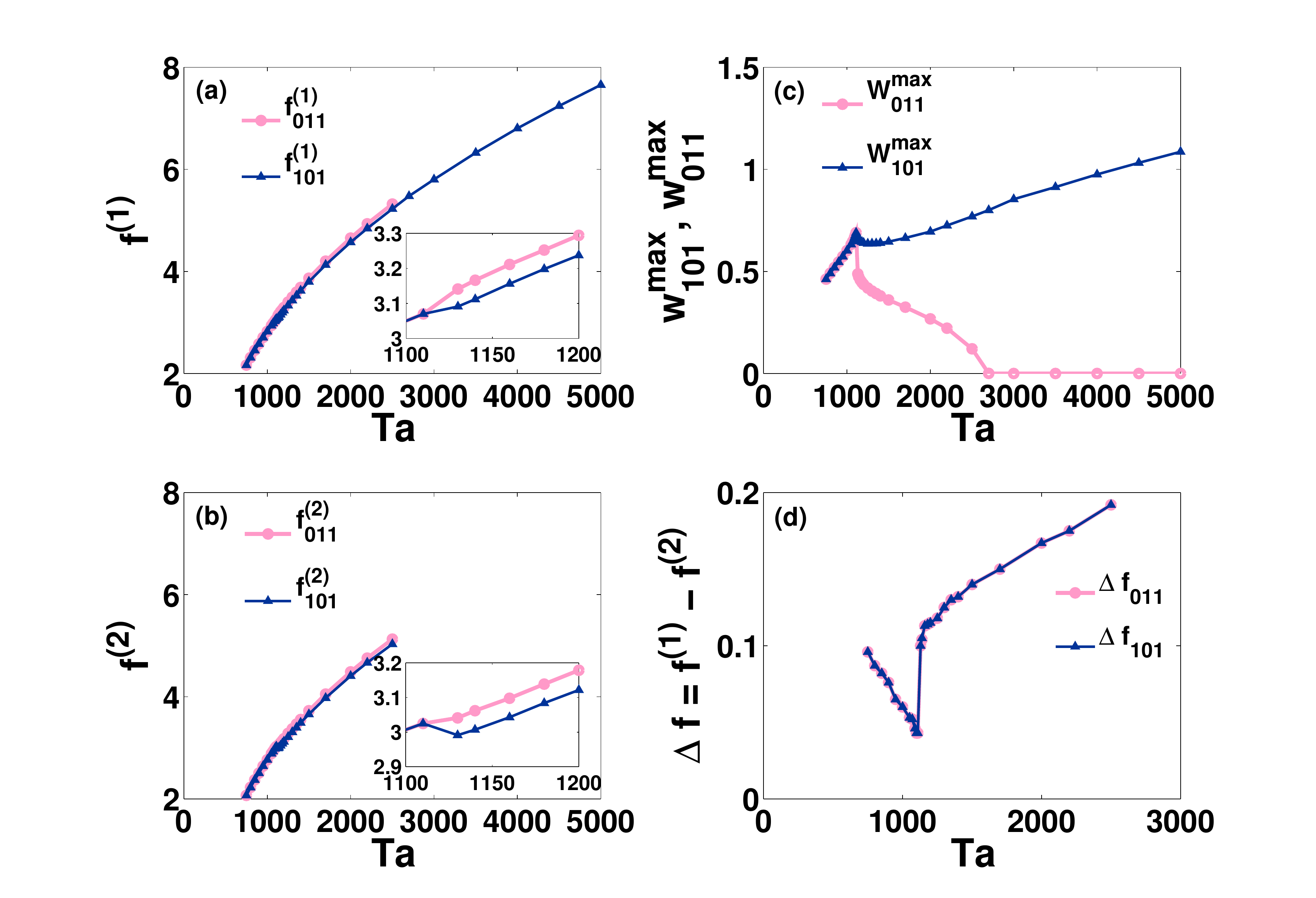}
\caption{(Color online) Bifurcations near the primary instability ($r = 1.005$) for $Pr =0.1$ and $\eta = 1$: (a) The variation of frequencies $f^{(1)}_{101}$ [blue (black) curve] and $f^{(1)}_{011}$ [pink (gray) curve] corresponding to the largest peak in the power spectra of the Fourier modes $W_{101}$ and $W_{011}$ with $Ta$, respectively. (b) The variation of frequencies $f^{(2)}_{101}$ [blue (black) curve] and $f^{(2)}_{011}$ [pink (gray) curve] corresponding to the second largest peak in the power spectra of the Fourier modes $W_{101}$ and $W_{011}$, respectively, with $Ta$. (c) The variation of the largest value of the modes $W_{011}$ [pink (gray) curve] and $W_{101}$ [blue (black) curve] with $Ta$. The maxima of the modes $W_{011}$ and $W_{101}$ are identical until $Ta = 1114$. The maximum of $W_{011}$ differs from the maximum of $W_{101}$ for $1125 \leq Ta < 2700$. (d) The variation of  $\Delta f_{101} = f^{(1)}_{101} - f^{(2)}_{101}$ [blue (black) curve] and $\Delta f_{011} = f^{(1)}_{011} - f^{(2)}_{011}$ [pink (gray) curve] with $Ta$.  $\Delta f_{101}$ and $\Delta f_{011}$ are always identical until $Ta = 2700$.  $f^{(1)}_{101}$ keeps increasing with $Ta$, but $f^{(1)}_{011}$,  $f^{(2)}_{101}$ and $f^{(2)}_{011}$ do not exist for $Ta \ge 2700$. The symbols  `$\triangle$' and `$\circ$' are data points computed from DNS for modes $W_{101}$ and $W_{011}$, respectively.} \label{bifurcation_sq}
\end{center}
\end{figure}

The largest and the second largest peaks in the power spectrum of Fourier mode $W_{101}$ (see, the third row of Fig.~\ref{psd_sq_box}) are located at frequencies  $f^{(1)}_{101}$ and  $f^{(2)}_{101}$, respectively. The two frequencies differ only slightly. Similarly, the power spectrum of the Fourier mode $W_{011}$ shows the largest and the second largest peaks at frequencies  $f^{(1)}_{011}$ and  $f^{(2)}_{011}$, respectively. The details of the bifurcation may be understood by studying the effects of rotation on these frequencies. The variations of  $f^{(1)}_{101}$ [blue (black) curve] and $f^{(1)}_{011}$ [pink (gray) curve] with $Ta$ are shown in Fig.~\ref{bifurcation_sq} (a) near the primary instability ($r = 1.005$) in a square box ($\eta = 1$). Both the frequencies for two modes remain exactly the same till $Ta \le 1114$, and become slightly different at $Ta \ge 1115$ [see the inset of Fig.~\ref{bifurcation_sq} (a)]. The maxima of the Nusselt number and the kinetic energy also showed peaks exactly at $Ta = 1114$ (Fig.~\ref{figure4}). This behavior continues till $Ta < 2700$. Only the frequency $f^{(1)}_{101}$ exists for  $Ta \ge 2700$, which then increases monotonically with $Ta$. Figure~\ref{bifurcation_sq} (b) shows the variation of frequencies $f^{(2)}_{101}$ and $f^{(2)}_{011}$ with $Ta$. They also remain initially equal, and then become unequal at $Ta \ge 1115$. Both the frequencies for two modes exist only for $Ta < 2700$. Figure~\ref{bifurcation_sq} (c) displays the variations of the maxima of modes $W_{101}$ [blue (black) curve] and $W_{011}$ [pink (gray) curve] with $Ta$. The maxima of modes $W_{101}$ and $W_{011}$ remain exactly the same until $Ta \le 1114$. They show a peak at $Ta = 1114$ and become unequal for $Ta \ge 1115$. The maximum of $W_{101}$ first decreases slightly, and then starts increasing  once again with $Ta$, while the maximum of $W_{011}$ decreases monotonically with increasing $Ta$. The mode  $W_{011}$ vanishes at $Ta = 2700$. The mode $W_{101}$ keeps increasing monotonically with $Ta$, as $Ta$ is raised further. Figure~\ref{bifurcation_sq}(d) shows the variation of the modulation frequency  $\Delta f_{101}$ $=$ $f^{(1)}_{101}$ $-$  $f^{(2)}_{101}$ of the mode $W_{101}$ and the modulation frequency $\Delta f_{011}$ $=$ $f^{(1)}_{011}$ $-$  $f^{(2)}_{011}$ of the mode $W_{011}$, respectively, with $Ta$. The modulation frequencies of both mode remain exactly the same, and siege to exist for $Ta \ge 2700$.

The patterns show a series of bifurcations. We observe a quasiperiodic competition between two sets of mutually perpendicular rolls (QPCR) of equal  maximum intensity at the primary instability for $750 \le Ta \le 1114$. A bifurcation occurs at $Ta = 1114.5 \pm 0.5$, and the flow becomes chaotic. The convection is chaotic in a narrow range of $Ta$ ($1115 \le Ta < 1125$). Another type of convective flow appears for $1125 \le Ta < 2700$. It now shows a quasiperiodic competition between two set of mutually perpendicular rolls of unequal  maximum intensity ($|W_{101}|$ $\neq$ $|W_{011}|$). A bifurcation from a three-dimensional quasiperiodic convection (QPCR) to 2D periodic convection occurs at $Ta = 2700$. The convection shows 2D standing waves instead of 3D  quasiperiodic cross-rolls at the primary instability for $Ta \ge 2700$. It is interesting to note that the variation of the maximum of $Nu-1$ and $E$ shows only signature of a bifurcation at $Ta = 1114$, when the flow becomes chaotic. 

\begin{figure}[h]
\begin{center}
\includegraphics[height=11 cm,width= 16 cm]{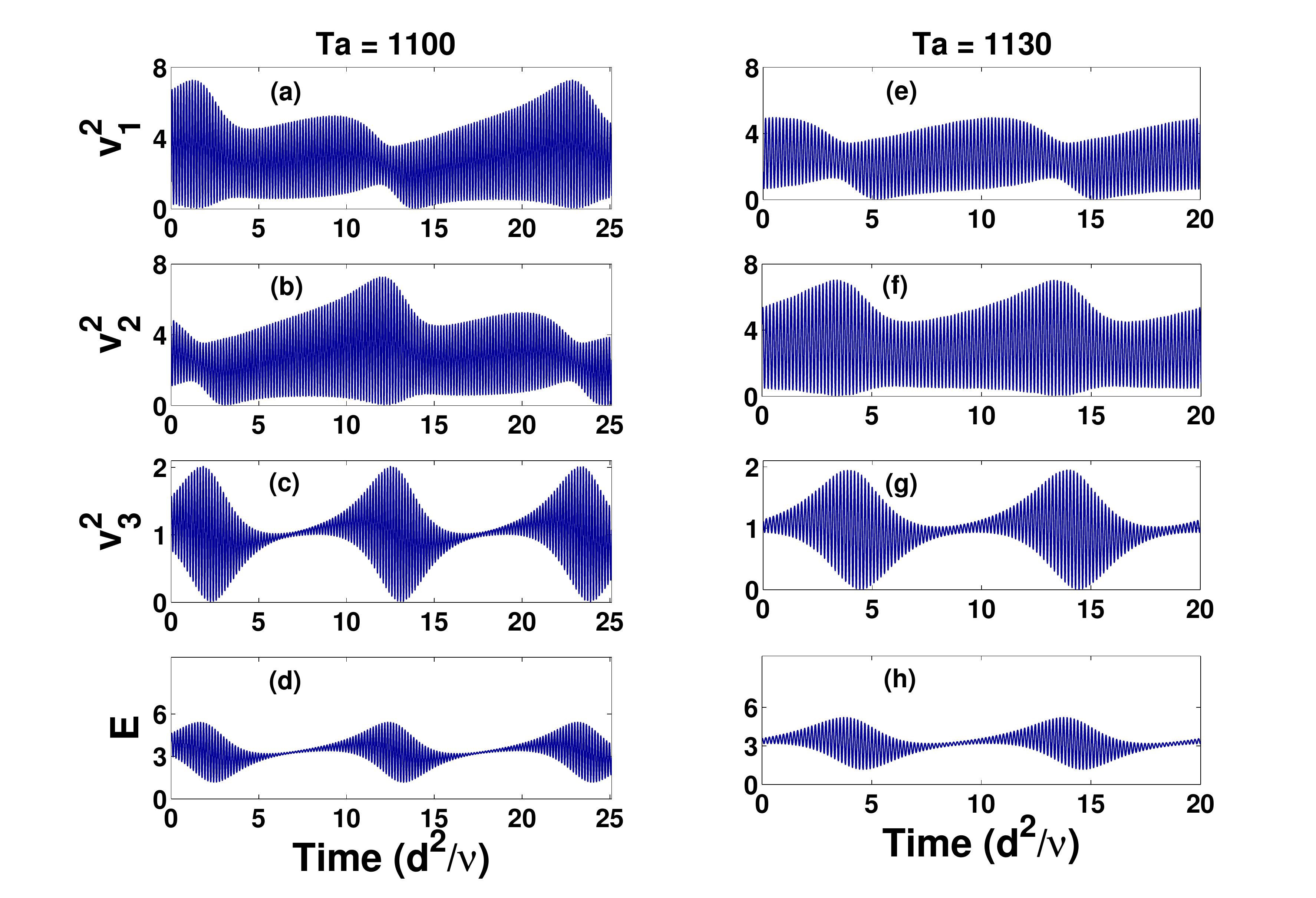}
\caption{(Color online) Temporal variation  of the box averaged quantities [in blue (black) color] just above the onset of convection ($r=1.005$, $Pr =0.1$) for $\eta = 1$. The left column displays the variation of (a) $v_1^2$, (b) $v_2^2$, (c) $v_3^2$ and (d) the kinetic energy $E = \frac{1}{2}(v_1^2 + v_2^2 + v_3^2)$ with time for $Ta =1100$. The right column shows the variation of (e) $v_1^2$, (f) $v_2^2$, (g) $v_3^2$ and (h) $E$ with time for $Ta = 1130$.} \label{E_Ta_1100_1130}
\end{center}
\end{figure}

\begin{figure}[h]
\begin{center}
\includegraphics[height= 10 cm,width= 14 cm]{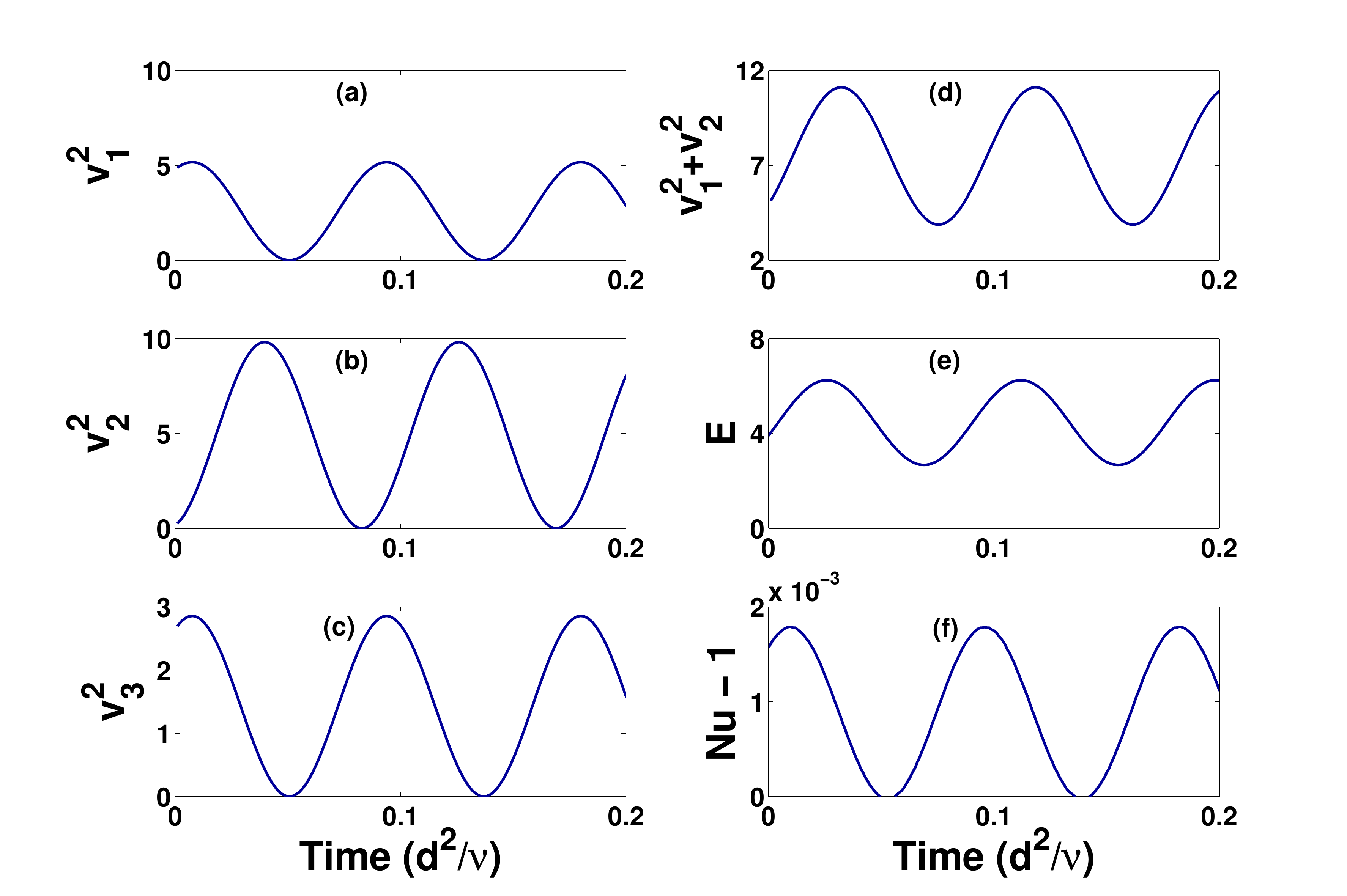}
\caption{(Color online) Temporal variation of the boxed averaged values of (a) $v_1^2$, (b) $v_2^2$, (c) $v_3^2$, (d) $v_1^2+ v_2^2$, (e) $E = (v_1^2 + v_2^2 + v_3^2)/2$, and (f) convective thermal flux ($Nu-1$) in blue (black) color just above the primary instability ($r=1.005$) for $Ta = 3000$, $Pr=0.1$ and $\eta =1$.} \label{E_Nu_Ta_3000}
\end{center}
\end{figure}

Figure~\ref{E_Ta_1100_1130} shows the temporal variations [blue (black) curve] of the spatially averaged values of  $v_1^2$ , $v_2^2$ , $v_3^2$ and the kinetic energy $E = (v_1^2 + v_2^2 + v_3^2)/2$ for $Ta = 1100$.  For any time dependent flow in a forced dissipative system like rotating RBC, the amount of energy injected in the system is not equal to the energy dissipated at every instant of time. However, the time averaged values of these quantities over a suitable period remain constant. Note the time averaged value of the kinetic energy  $\bar{E}$ is a finite positive constant, and therefore the same for $\frac{dE}{dt}$ is equal to zero. As $\frac{dE}{dt}$ fluctuates in time, the energy $E(t)$ also varies from its average value.  The energy $E$ and the convective heat flux ($Nu - 1$) vary quasiperiodically in time in this case.  The period of amplitude modulations of $v_1^2$ [Fig.~\ref{E_Ta_1100_1130}(a)] and $v_2^2$ [Fig.~\ref{E_Ta_1100_1130}(b)] is equal to twice the period of amplitude modulation for $v_3^2$ [Fig.~\ref{E_Ta_1100_1130}(c)] for $Ta =1100$.  Although the period of amplitude modulations for $v_1^2$ and $v_2^2$ is the same, they are out of phase. The period of amplitude modulation for $E$ and $Nu$ (not shown here) is similar to that of $v_3^2$. The second column of Fig.~\ref{E_Ta_1100_1130} displays similar plots for $Ta = 1130$. The period of amplitude modulation is the same for all the quantities $v_1^2$, $v_2^2$, $v_3^2$, and $E$ in this case. The amplitude modulations for $v_1^2$ and $v_2^2$ are again out of phase. 

\begin{figure}[h]
\begin{center}
\includegraphics[height=11 cm,width=12 cm ]{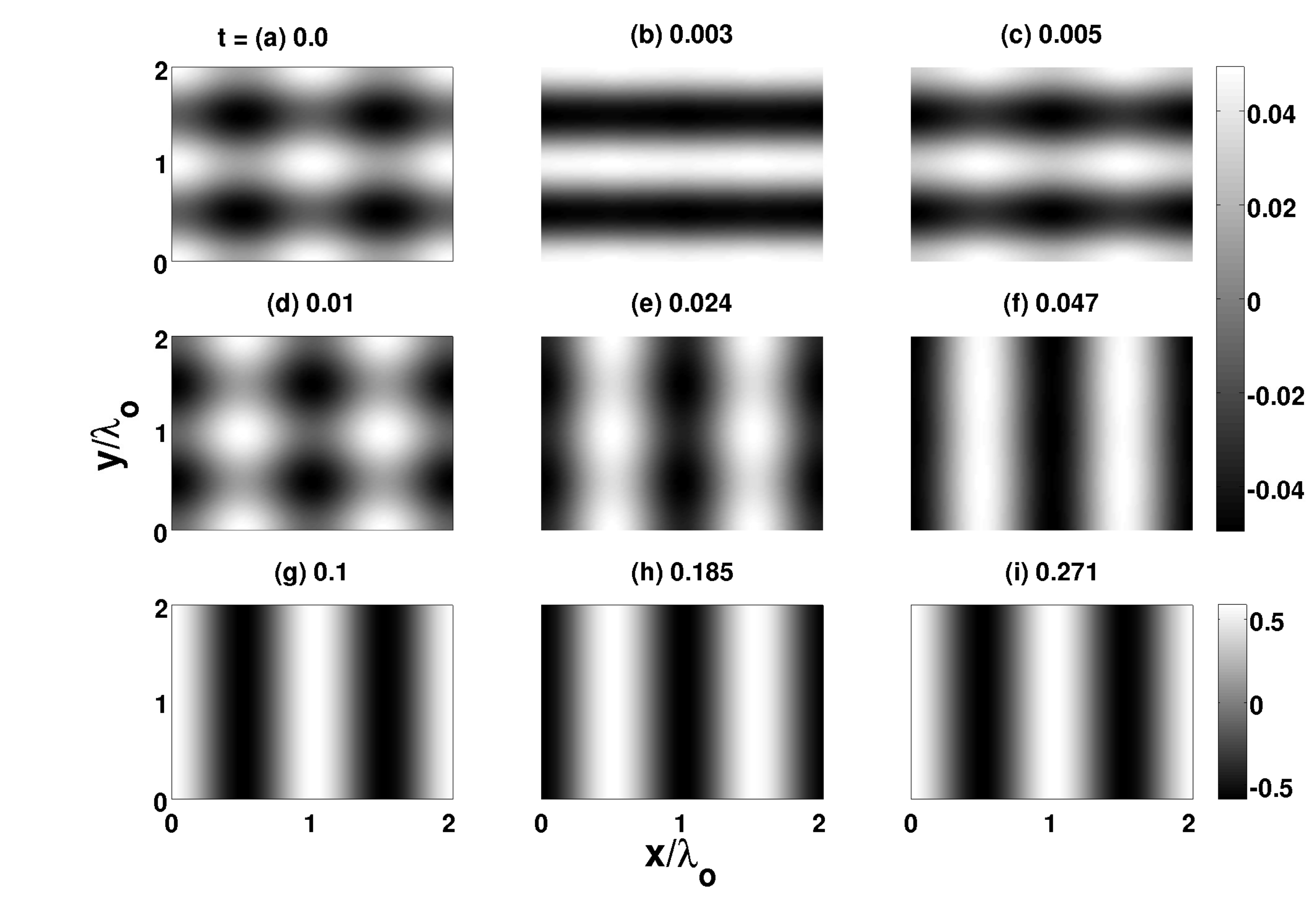}
\caption{Transition from three-dimensional (3D) convection to two-dimensional (2D) convection ($Pr = 0.1$) near onset ($r= 1.005$) in a square box ($\eta =1$) as the Taylor number $Ta$ is varied. Mid-plane temperature contour plots at different instants [(a)\---(f)] of 3D quasiperiodic patterns for $Ta = 2500$ display a competition between two sets of mutually perpendicular rolls. Different phases [(g)\---(i)] of a 2D standing wave for $Ta = 3000$ are shown in the last row.} \label{contour_3d_to_2d}
\end{center}
\end{figure}

Figure~\ref{E_Nu_Ta_3000} displays the temporal variation of (a) $v_1^2$, (b) $v_2^2$, (c) $v_3^2$, (d) $v_1^2 + v_2^2$, (e) $E$ and (f) $Nu-1$ for $Ta = 3000$. All the fields become periodic. Notice that all the velocity components $v_1$, $v_2$ and $v_3$ are nonzero. A bifurcation from 3D quasiperiodic convection to 2D periodic convection is displayed in Fig.~\ref{contour_3d_to_2d}. The upper and middle rows [see Fig.~\ref{contour_3d_to_2d}(a)\---(f)] show the contour plots of the temperature field for six different instants of time for $Ta = 2500$, while the lower row [Fig.~\ref{contour_3d_to_2d}(g)\---(i)] displays the convective patterns at different instants of time for $Ta = 3000$. This shows a bifurcation from 3D convection to 2D convection, which occurs at $Ta=2700$.  
A set of parallel rolls in the presence of Coriolis force always has all the three velocity components nonzero.  This can be easily seen from the expressions for the vertical velocity $v_3$ and vertical vorticity $\omega_3$ ($= \hat{\boldsymbol{z}}\boldsymbol{\cdot \omega} = \hat{\boldsymbol{z}}\boldsymbol{\cdot \nabla \times v}$) for a set of straight rolls parallel to the $y$ axis. Considering only the critical modes, the fields $v_3$ and $\omega_3$ may be written as:
\begin{equation}
v_3 (x, z, t) = W_{101} (t) \cos{(k_c x)} \sin{(\pi z)}, ~~\ ~~\  \omega_3 (x, z, t) = \zeta_{101} (t) \cos{(k_c x)} \cos{(\pi z)}. \label{linear_modes}
\end{equation}
The horizontal velocities $v_1$ and $v_2$ may be obtained using the following relations:
\begin{equation}
\nabla_H^2 v_1 = -\partial_{xz} v_3 - \partial_y \omega_{3}, ~~\ ~~\  \nabla_H^2 v_2 = -\partial_{yz} v_3 + \partial_x \omega_{3}. \label{v1_v2}
\end{equation}
The expressions for the horizontal velocities $v_1$ and $v_2$ are given by,
\begin{eqnarray}
v_1 (x, z, t) &=& -(\pi/k_c) W_{101} (t) \sin{(k_c x)} \cos{(\pi z)},\nonumber\\
v_2 (x, z, t) &=& (1/k_c) \zeta_{101} (t) \sin{(k_c x)} \cos{(\pi z)}. 
\end{eqnarray}
This clearly shows that all the velocity components are nonzero, if the vertical vorticity is nonzero. In fact, it is the velocity $v_2$ that breaks the mirror symmetry of the steady $2D$ rolls parallel to the $y$ axis. 

\begin{figure}[h]
\begin{center}
\includegraphics[height= 10 cm,width= 14 cm]{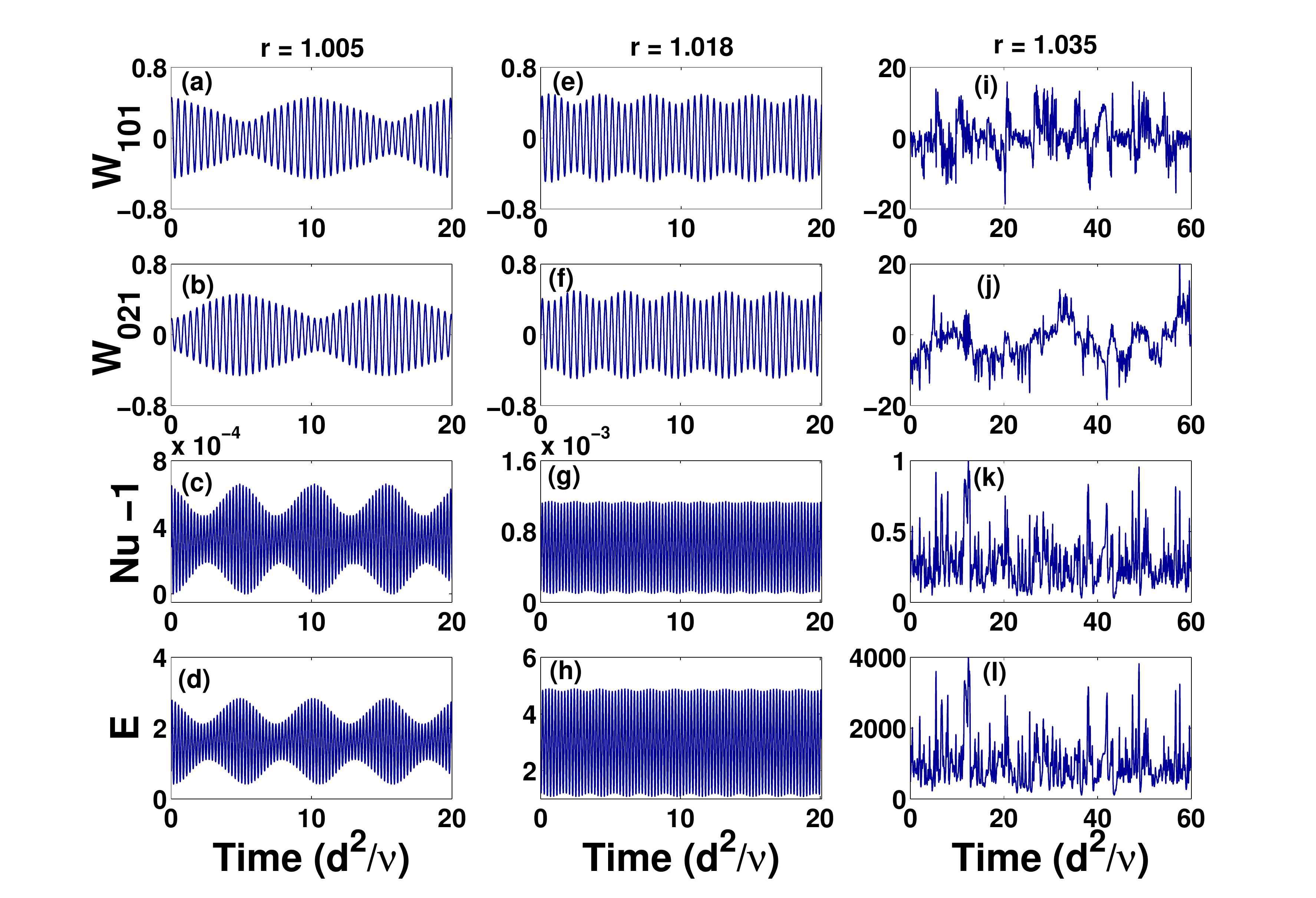}
\caption{(Color online) Temporal variation of the Fourier modes $W_{101}$, $W_{021}$, the convective heat flux $Nu-1$, and  the average kinetic energy $E$ with dimensionless time for Taylor number $Ta = 750$, $Pr = 0.1$, and $\eta = 2$. The blue (black) curves in the first column [(a)\---(d)], the second column [(e)\---(h)], and the third column [(i)\---(l)] are for the reduced Rayleigh number $r = 1.005$, $1.018$ and $1.035$, respectively.} \label{l2}
\end{center}
\end{figure}
\begin{figure}[h]
\begin{center}
\includegraphics[height=14 cm,width= 11.5 cm]{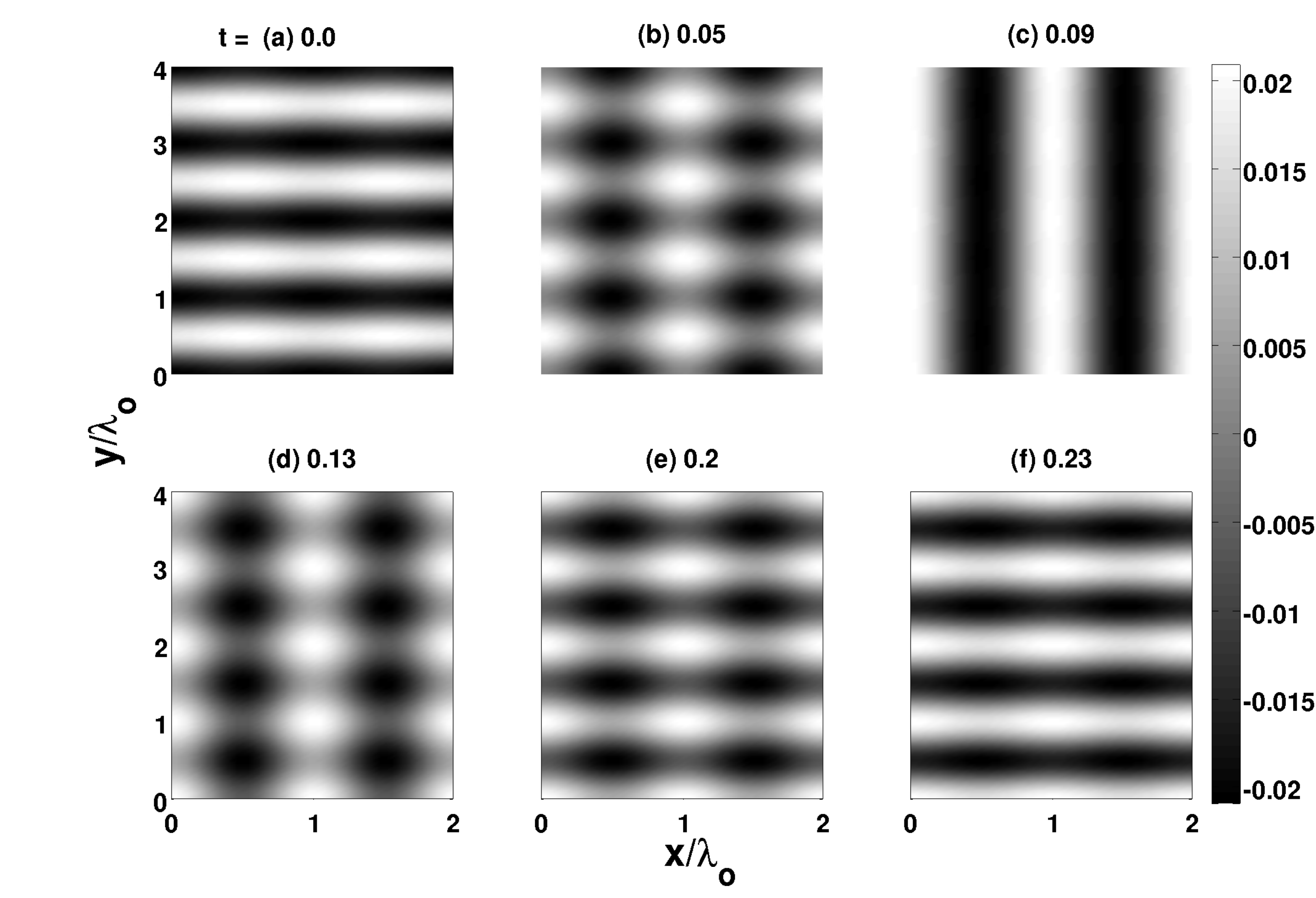}
\caption{The contour plots of the temperature field near the onset ($r =1.005$) of convection at $z=0.5$ computed from DNS for $\eta =2$ and $Ta = 750$. The patterns are shown in a region four times the size of the simulation box for clarity.} \label{contour_eta_2}
\end{center}
\end{figure}

\begin{figure}[h]
\begin{center}
\includegraphics[height=11 cm,width= 14 cm]{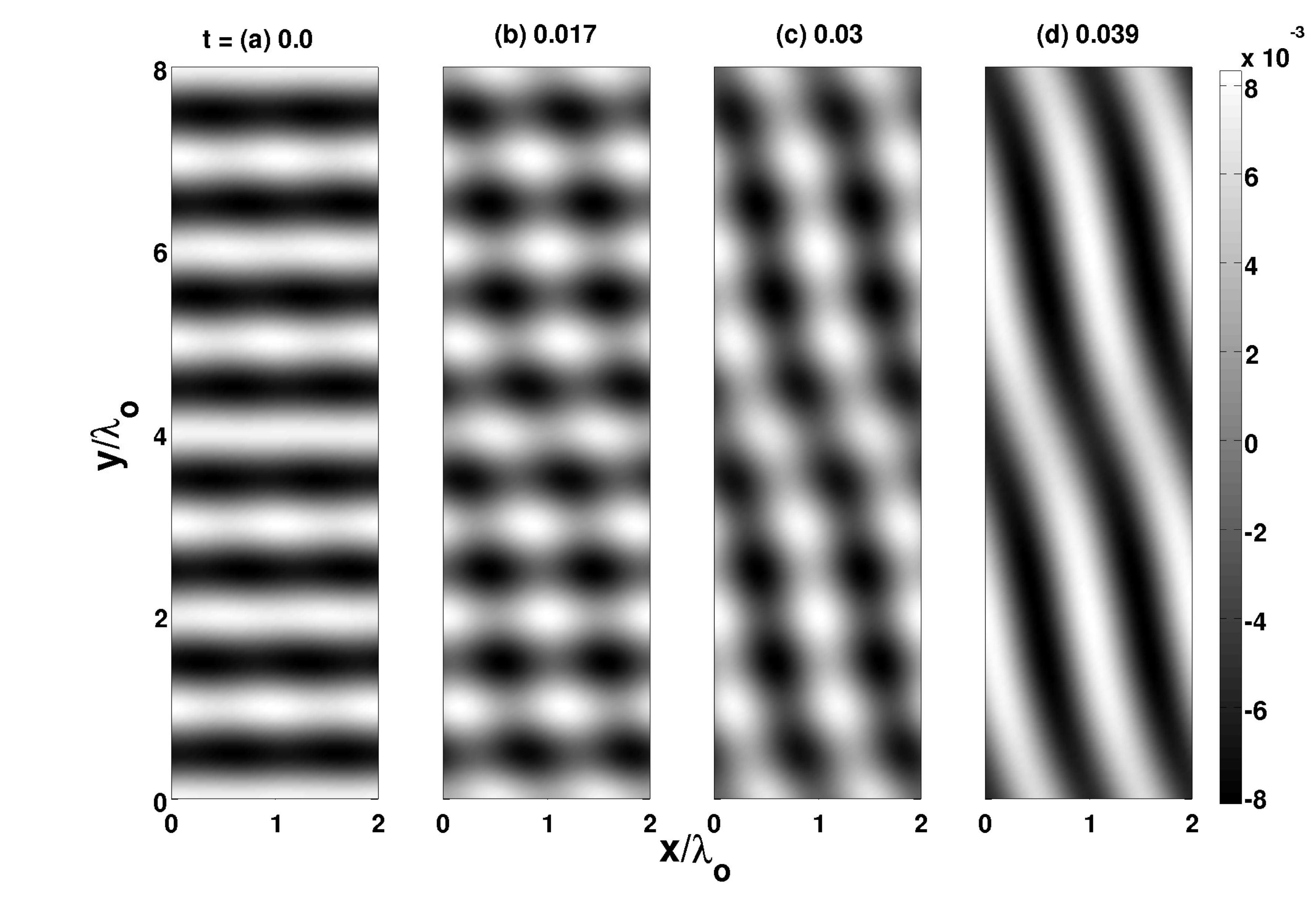}
\caption{Contour plots (a)\---(d) for the temperature field in the mid-plane ($z=0.5$) for four different time instants for the horizontal aspect ratio $\eta = 4$, $Ta = 750$ and $r =1.005$.} \label{contour_eta_4}
\end{center}
\end{figure}

\begin{figure}[h]
\begin{center}
\includegraphics[height=10 cm,width= 14 cm]{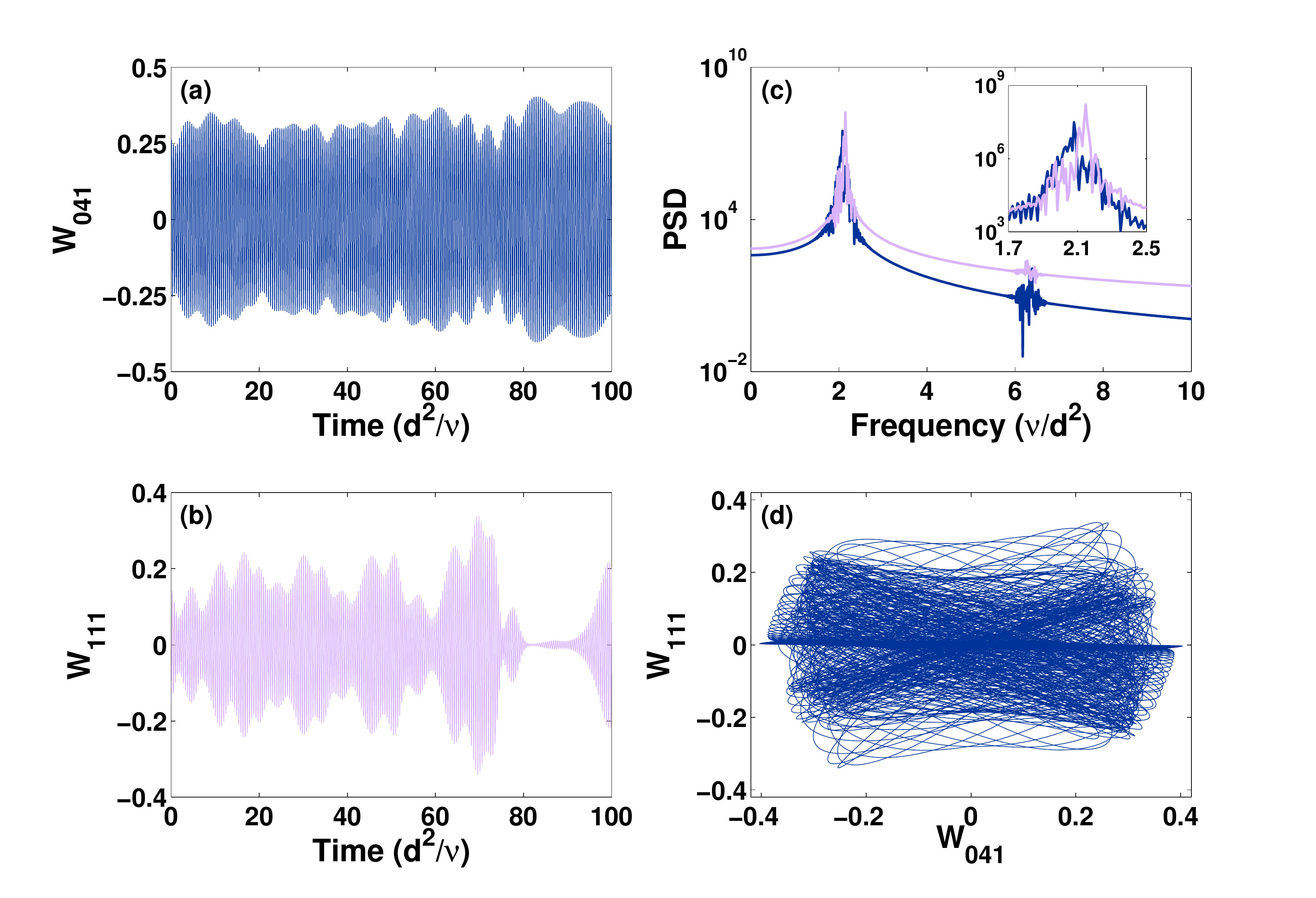}
\caption{(Color online) Details of Fourier modes $W_{041}$ [blue (black) curve] and $W_{111}$ [purple (gray) curve] at the instability onset ($r =1.005$) for $\eta = 4$ and $Ta = 750$. The temporal variation of (a) $W_{041}$ and (b) $W_{111}$ show chaotic nature of the signal. The power spectra of these two modes are shown in (c), and phase portrait in the plane $W_{041} - W_{111}$ is displayed in (d).} \label{signal_details_eta_4}
\end{center}
\end{figure}

\subsection{Patterns in a rectangular box ($\eta \ge 2$)}
We now present the results of our simulations in a rectangular box ($\eta = L_y/L_x \ge 2$). We begin with the results of DNS for $\eta =2$. The temporal variations of the two largest Fourier modes $W_{101}$ and $W_{021}$, the convective heat flux ($Nu-1$), and the kinetic energy $E$ near the instability onset are displayed for $Ta = 750$ and $Pr =0.1$ in Fig.~\ref{l2}. The convection is temporally quasiperiodic at the onset (see the first column of Fig.~\ref{l2} for $r=1.005$). The amplitude modulations of the Fourier modes $W_{101}$ [Fig.~\ref{l2} (a)] and $W_{021}$ [Fig.~\ref{l2} (b)] are out of phase. However, the convective heat flux ($Nu-1$) [Fig.~\ref{l2} (c)] across the fluid layer and the kinetic energy $E$ [Fig.~\ref{l2} (d)] are in phase. The mid-plane contour plots of the temperature field at the instability onset for the simulation box with $\eta =2$ are displayed in Fig.~\ref{contour_eta_2}. The convective patterns (Fig.~\ref{contour_eta_2}) are  quasiperiodically oscillating cross-rolls (QPCR), as observed in a square box (Fig.~\ref{contour_Ta_750}). The period of amplitude modulation decreases with increase in $r$ [see Fig.~\ref{l2}(e) - (h)]. As $r$ is raised to a value just $3\%$ above the threshold for oscillatory convection, the flow becomes chaotic in time (see the third column of Fig.~\ref{l2}). The chaotic behavior was also predicted by Dawes~\cite{dawes_physicaD_2000} at the onset for much higher values of $Ta$ ($\ge 9 \times 10^4$) in a fluid with $Pr =0.1$.  We observe the chaotic flow at a much smaller value of $Ta$, and it occurs as a secondary instability.  

We observe an interesting change in the fluid pattern dynamics at the instability onset for $\eta =4$ (see Fig.~\ref{contour_eta_4}). Two sets of rolls inclined at an angle of $\phi$ $=$ $\pi/2$ $-$ $\arctan{(k_y/k_x)}$ compete with each other. The patterns vary in time chaotically. We observe a gradual transition from one set of wavy rolls oriented along the $x$ axis to another set of rolls oriented at an angle of approximately $76^{\circ}$ with the $x$ axis. We observe rhombic patterns during the transition. Figure~\ref{signal_details_eta_4} gives the details of the two largest Fourier modes $W_{041}$ and $W_{111}$ for patterns shown in Fig.~\ref{contour_eta_4}. The temporal variation of both the modes [see, Fig.~\ref{signal_details_eta_4}(a), (b)] shows chaotic behavior. The time average of both the modes is zero. The phase between the maximum or minimum of these two modes varies  with time. The power spectra  of $W_{041}$ [blue (black) curve] and $W_{111}$ [purple (gray) curve] are shown in Fig.~\ref{signal_details_eta_4} (c). The details of the first largest peak [see the inset of Fig.~\ref{signal_details_eta_4}(c)] show a narrow band of frequencies instead of a few discrete ones. The largest peak in the power spectra for both the modes $W_{041}$ and $W_{111}$ appear at slightly different frequencies. The temporal evolution of any trajectory in the $W_{041} - W_{111}$ plane shows a chaotic behavior. 

The convective patterns shown in Fig.~\ref{contour_eta_4} have similarity with K\"{u}ppers-Lortz instability,~\cite{kuppers_lortz} where the change in orientation of rolls are determined by noise. The change in orientation of rolls occur here due to the presence of the wavy modes (e.g., $W_{111}$) in a longer container. The waves along the roll axis are easily excited in low-Prandtl-number fluids. The longer box size allows the distortion of fields easily even at slow rotation rates. The nonlinear interactions of these perturbations with the critical modes lead to the excitation of more than two independent frequencies. This causes temporally chaotic patterns at the onset of convection. The oblique wavy rolls appear as soon as the mode $W_{041}$ vanishes, but the value of $W_{111}$ remains relatively large. The wavy rolls are oriented along the $x$ axis when the mode $W_{111}$ is vanishingly small. We observe rhombic patterns when both the modes are nonzero. The angle between the two sets of wavy rolls is fixed, and is equal to $\phi = \frac{\pi}{2} - \arctan{(\eta^{-1})}$, although the dynamics is chaotic. The convective patterns observed at the primary instability are similar in other larger containers ($4 \le \eta \le 10$).

\section{A Low-dimensional model for square box}
We now construct a low-dimensional model for the Rayleigh-B\'{e}nard convection with rotation in a square box ($\eta = 1$). We first recast the hydrodynamic equations in a convenient form for this purpose. We operate by curl twice on Eq.~\ref{eq:momentum} and use the equation of continuity (Eq.~\ref{eq:continuity}). The equations for the vertical velocity $v_3$ and the vertical vorticity $\omega_3$ are then given by,
\begin{eqnarray}
\partial_{t}(\nabla^{2} v_{3}) &=& \nabla^{4} v_{3} + Ra{\nabla^{2}_{H}}\theta - 
\sqrt{Ta}\partial_{z}\omega_{3} - \hat{{\bf z}} {\bf \cdot}[\bf{\nabla\times}\{ ({\boldmath{\omega}}{\bf{\cdot\nabla}})
\bf{v}-(\bf{v\cdot\nabla}){\boldmath{\omega}})\}],\label{velocity}\\ 
\partial_{t}\omega_{3} &=& \nabla^{2}\omega_{3} + \sqrt{Ta} \partial_{z}v_{3} 
 + [({\bf{\omega \cdot\nabla}})v_3 - ({\bf{v\cdot\nabla}})\omega_3].
\label{vorticity}
\end{eqnarray} 
The equation (Eq.~\ref{eq:theta}) of temperature field remains unchanged. The spatial dependence of all the fields is expanded in terms of Fourier series with appropriate horizontal and vertical boundary conditions. The Fourier modes found to carry higher energy in the direct numerical simulations (DNS) are retained for the construction of the low-dimensional model. The vertical velocity $ v_{3}$, the vertical vorticity $\omega_{3}$ and the temperature $\theta$ fields are expressed as:
\begin{equation}
v_3 (x, y, z, t)= [W_{101}(t)\cos{kx}+W_{011}(t)\cos{ky}]\sin{\pi z}
+W_{112}(t)\cos{kx} \cos{ky} \sin{2\pi z},
\end{equation}
\begin{eqnarray}
\omega_3 (x, y, z, t) &=& [Z_{101}(t)\cos{kx}+Z_{011}(t)\cos{ky}]\cos{\pi z}
+Z_{110}(t)\cos{kx}\cos{ky}\nonumber\\
&+& Z_{112}(t)\cos{kx}\cos{ky}\cos{2\pi z}+Z_{200}(t)\cos{2kx}+Z_{020}(t)\cos{2ky}\nonumber\\ 
&+& Z_{130}(t)\sin{kx}\sin{3ky} +Z_{310}(t)\sin{3kx}\sin{ky},
\end{eqnarray}
\begin{eqnarray}
\theta_3 (x, y, z, t) &=& [\theta_{101}(t)\cos{kx}+\theta_{011}(t)\cos{ky}]\sin{\pi z}
+\theta_{112}(t)\cos{kx}\cos{ky}\sin{2\pi z}\nonumber\\
&+&\theta_{002}(t)\sin{2\pi z}. 
\end{eqnarray}

Projecting the hydrodynamical equations (Eqs.~\ref{velocity}-\ref{vorticity}) and the equation for the temperature field (Eq.~\ref{eq:theta}) onto these modes, we arrive at a set of fifteen ordinary nonlinear differential equations. Fourier modes $W_{112}, Z_{112}, Z_{130}, Z_{310}, \theta_{112}, \theta_{002}$, which decay linearly much faster than the other modes, are then adiabatically eliminated. This leads to a nine-mode model given by:
\begin{eqnarray}
\dot{X_1} &=& \frac{1}{\alpha} \left [Rak^{2}{Y_1} -\alpha^{2}{X_1} - \frac{\pi}{2}G{Z_2} -\pi \sqrt{Ta}{Z_1}\right. \nonumber\\
&-&  \left. \frac{\pi}{4}(\alpha{X_2} - \pi \beta \sqrt{Ta}{Z_2}) \left(\frac{\gamma X_1X_2+\delta Z_1Z_2+\pi Pr Ra k^2(X_1Y_2+X_2Y_1)}{4\pi^2Ta-2 Ra k^2+ \chi +2\pi^2\sqrt{Ta}(S_1+S_2)}\right) \right], \label{model_eq1a}
\end{eqnarray}
\begin{eqnarray}
\dot{X_2} &=& \frac{1}{\alpha} \left [Rak^{2}{Y_2} -\alpha^{2}{X_2} - \frac{\pi}{2}G{Z_1} -\pi \sqrt{Ta}{Z_2}\right. \nonumber\\
&-&  \left. \frac{\pi}{4}(\alpha{X_1} - \pi \beta \sqrt{Ta}  {Z_1}) \left(\frac{\gamma X_1X_2+\delta Z_1Z_2+\pi Pr Ra k^2(X_1Y_2+X_2Y_1)}{4\pi^2Ta-2 Ra k^2+ \chi +2\pi^2\sqrt{Ta}(S_1+S_2)}\right) \right], \label{model_eq1b}
\end{eqnarray}
\begin{eqnarray}
\dot{Y_1} &=& \frac{1}{Pr}({X_1}-\alpha{Y_1}) -\frac{Pr}{8}(X_1Y_1+X_2Y_2){X_1} - \frac{\pi^2\beta Pr}{16} (X_1Y_2+ X_2Y_1){X_2}\nonumber\\
&-& \frac{\pi\beta}{8} \left(\frac{\gamma X_1X_2+\delta Z_1Z_2+\pi Pr Ra k^2(X_1Y_2+X_2Y_1)}{4\pi^2Ta-2 Ra k^2+ \chi +2\pi^2\sqrt{Ta}(S_1+S_2)}\right){X_2}, \label{model_eq2a}
\end{eqnarray}
\begin{eqnarray}
\dot{Y_2} &=& \frac{1}{Pr}({X_2}-\alpha{Y_2}) -\frac{Pr}{8}(X_1Y_1+X_2Y_2){X_2} - \frac{\pi^2\beta Pr}{16} (X_1Y_2+ X_2Y_1){X_1}\nonumber\\
&-& \frac{\pi\beta}{8} \left(\frac{\gamma X_1X_2+\delta Z_1Z_2+\pi Pr Ra k^2(X_1Y_2+X_2Y_1)}{4\pi^2Ta-2 Ra k^2+ \chi +2\pi^2\sqrt{Ta}(S_1+S_2)}\right){X_1}, \label{model_eq2b}
\end{eqnarray}
\begin{eqnarray}
\dot{Z_1} &=& \pi \sqrt{Ta}{X_1} - \alpha{Z_1} - \frac{\pi}{2} X_1S_1 \nonumber\\ 
&-& \frac{\pi}{4}({Z_2}+\pi \beta \sqrt{Ta}{X_2})\left(\frac{\gamma X_1X_2+\delta Z_1Z_2+\pi Pr Ra k^2(X_1Y_2+X_2Y_1)}{4\pi^2Ta-2 Ra k^2+ \chi +2\pi^2\sqrt{Ta}(S_1+S_2)}\right), \label{model_eq3a}
\end{eqnarray}
\begin{eqnarray}
\dot{Z_2} &=& \pi \sqrt{Ta}{X_2} - \alpha{Z_2} - \frac{\pi}{2} X_2S_2 \nonumber\\ 
&-& \frac{\pi}{4}({Z_1}+\pi \beta \sqrt{Ta}{X_1})\left(\frac{\gamma X_1X_2+\delta Z_1Z_2+\pi Pr Ra k^2(X_1Y_2+X_2Y_1)}{4\pi^2Ta-2 Ra k^2+ \chi +2\pi^2\sqrt{Ta}(S_1+S_2)}\right), \label{model_eq3b}
\end{eqnarray}
\begin{eqnarray}
\dot{S_1} &=& \pi X_1Z_1 - 4k^2{S_1} - \frac {1}{200k^2}G^2{S_1}\nonumber\\
&+&\frac{\pi^2 \beta}{2}\sqrt{Ta}\left(\frac{\gamma X_1X_2+\delta Z_1Z_2+\pi Pr Ra k^2(X_1Y_2+X_2Y_1)}{4\pi^2Ta-2 Ra k^2+ \chi +2\pi^2\sqrt{Ta}(S_1+S_2)}\right)^2,\label{model_eq4a}
\end{eqnarray}
\begin{eqnarray}
\dot{S_2} &=& \pi X_2Z_2 - 4k^2{S_2} - \frac {1}{200k^2}G^2{S_2}\nonumber\\
&+&\frac{\pi^2 \beta}{2}\sqrt{Ta}\left(\frac{\gamma X_1X_2+\delta Z_1Z_2+\pi Pr Ra k^2(X_1Y_2+X_2Y_1)}{4\pi^2Ta-2 Ra k^2+ \chi +2\pi^2\sqrt{Ta}(S_1+S_2)}\right)^2,\label{model_eq4b}
\end{eqnarray}
\begin{equation}
\dot{G} = \pi(X_1Z_2 + X_2Z_1)- 2k^2G + \frac{3}{800k^2}G(S_1^2+S_2^2),  \label{model_eq5}
\end{equation}
where
$X_1 = W_{101}$, $X_2 = W_{011}$, $Y_1 = \theta_{101}$, $Y_2 = \theta_{011}$, $Z_1 = Z_{101}$, $Z_2 = Z_{011}$, $S_1 =Z_{200}$, $S_2 = Z_{020}$,  $G$ $=$ $Z_{110}$, $\alpha$ $=$ $(k^2+\pi^2)$ , $\beta$ $=$  $(k^2 +2\pi^2)^{-1}$, $\gamma$ $=$ $8\pi^5+4\pi k^4 +12\pi^3 k^2$, $\delta$ $=$ $(8\pi^3 +4\pi k^2)$, and $\chi$ $=$ $64\pi^6+8k^6+48\pi^2k^4+96\pi^4k^2$. 

\begin{figure}[h]
\begin{center}
\includegraphics[height=10 cm,width= 14 cm]{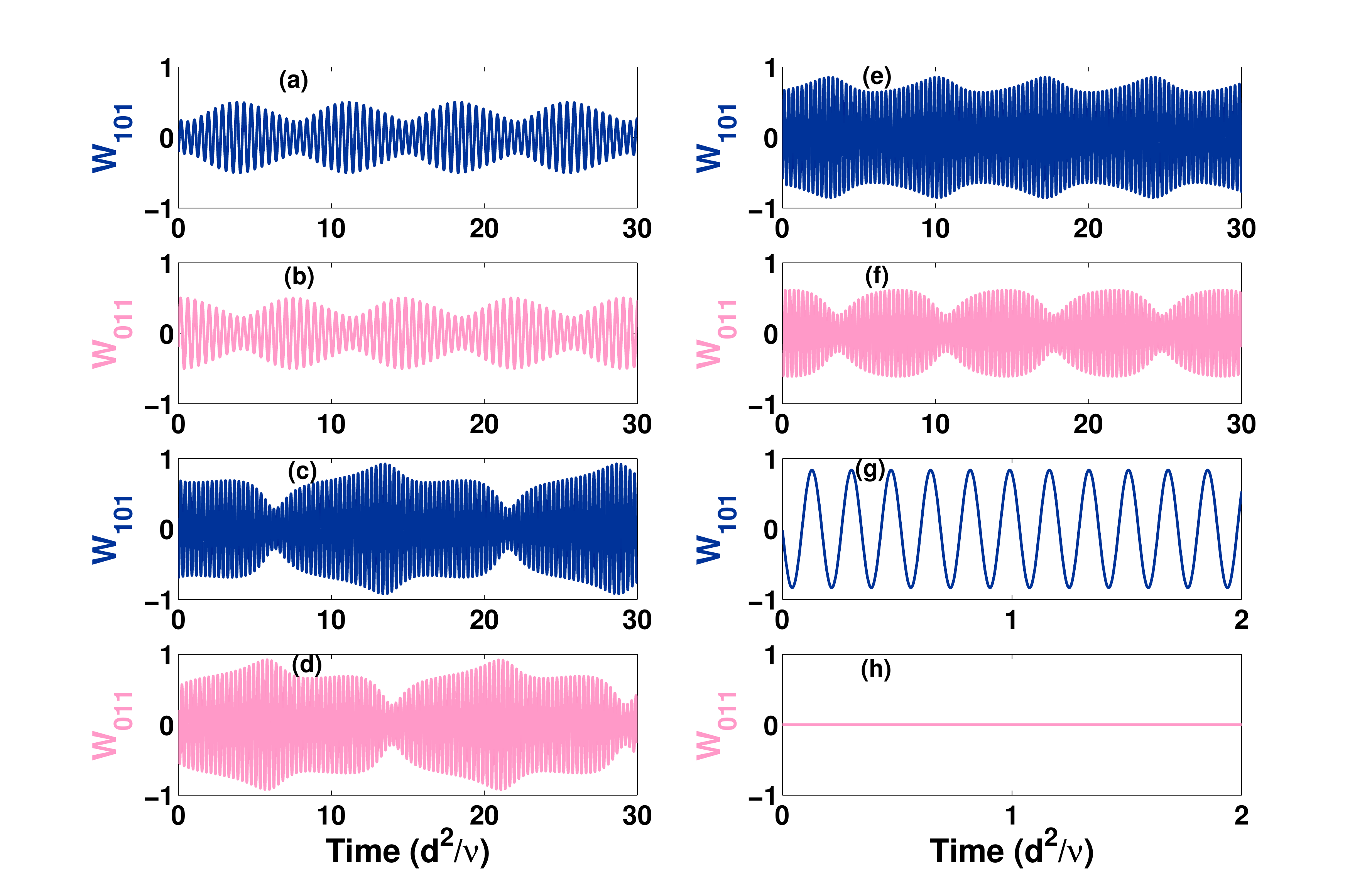}
\caption{(Color online) Temporal variation of the Fourier modes $W_{101}$ [blue (black) curve] and $W_{011}$ [pink (gray) curve] just above the  onset ($r = 1.005$), as computed from the low-dimensional model ($\eta = 1$): (a) $W_{101}$ \& (b) $W_{011}$ for $Ta=750$, (c) $W_{101}$ \& (d) $W_{011}$ for $Ta=1290$,  (e) $W_{101}$ \& (f) $W_{011}$ for $Ta=1320$,  and (g) $W_{101}$ \& (h) $W_{011}$ for $Ta=3000$.} \label{modes_model}
\end{center}
\end{figure}

\begin{figure}[h]
\begin{center}
\includegraphics[height=10 cm,width= 14 cm]{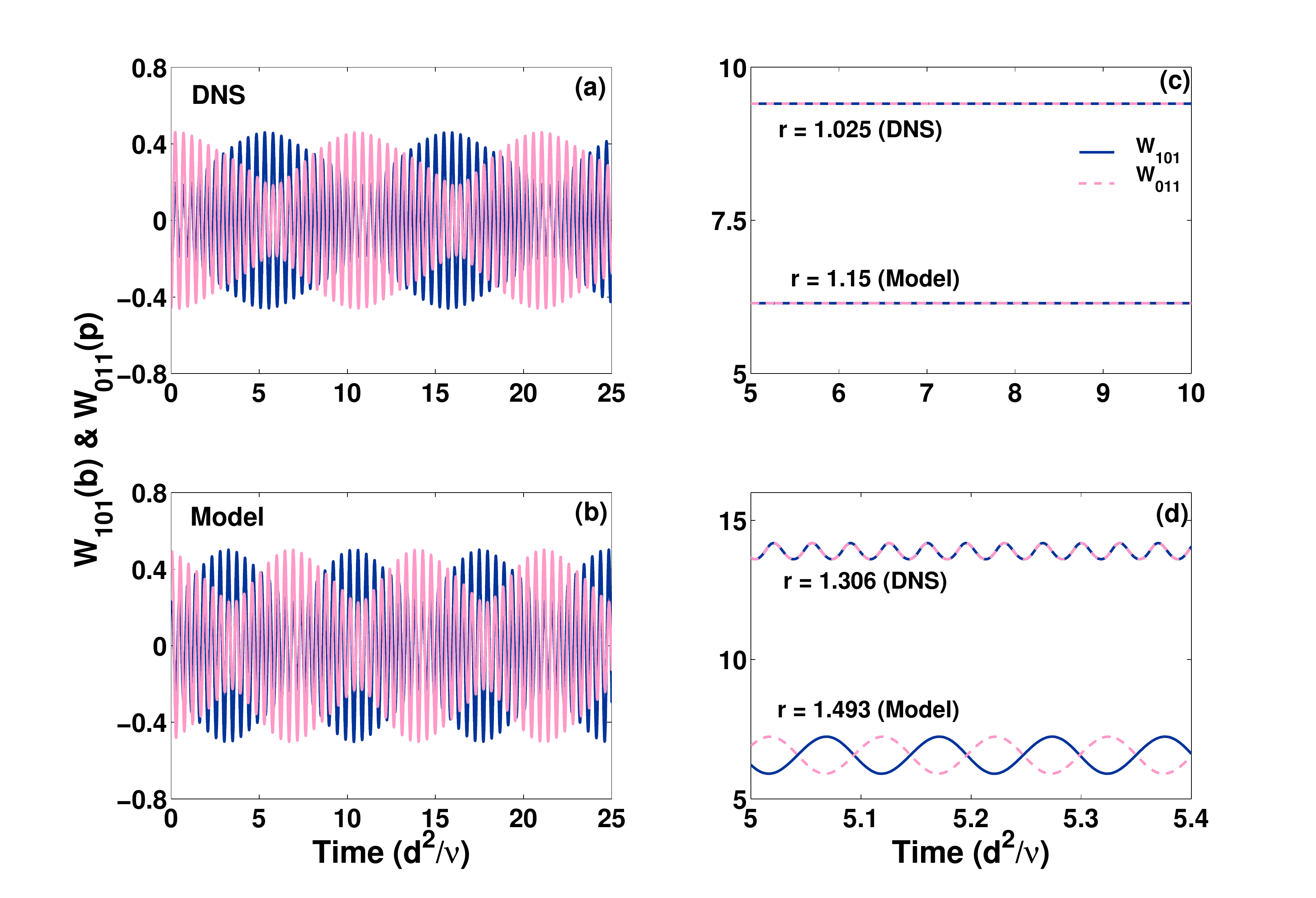}
\caption{(Color online) Comparison of the results obtained from DNS and the model for $Pr = 0.1$, $Ta = 750$, and $\eta = 1$. The first column shows the temporal variation of Fourier modes $W_{101}$ [blue (black) curve] and $W_{011}$ [pink (gray) curve] near the onset of convection ($r =1.005$) computed from (a) DNS and (b) the model. The second column displays (c) the values of $W_{101}$ and $W_{011}$ for stationary square pattern near the secondary instability ($r = 1.025$ in DNS and $r = 1.15$ in the model) and (d) the temporal variation of $W_{101}$ and $W_{011}$ near tertiary instability ($r = 1.306$ in DNS and $r = 1.493$ in the model).} \label{model_dns}
\end{center}
\end{figure}

Figure~\ref{modes_model} shows the temporal variation of the leading modes at four different rotation rates-- $Ta=750$, $Ta=1290$, $Ta=1320$, and $Ta=3000$ obtained from the low-dimensional model (Eqs.~\ref{model_eq1a}-\ref{model_eq5}). The results computed from the model are in good agreement with those obtained from DNS at the primary instability, and in qualitative agreement with those at secondary and tertiary instabilities. At $Ta = 750$, both the modes $W_{101}$ and $W_{011}$ are found to oscillate quasiperiodically just above  the onset of oscillatory convection. This behavior continues to exist until $Ta = 1301$. The primary instability is temporally chaotic in a small range of the Taylor numbers $1301 < Ta < 1311$. At $Ta = 1311$, a bifurcation occurs and both the leading modes $W_{101}$ and $W_{011}$ show quasiperiodic temporal variation with different modulation amplitudes and frequencies. The bifurcation from 3D quasiperiodic behavior into 2D periodic behavior occurs at $Ta = 3000$. One of the roll-modes becomes zero and the other oscillates with zero mean. The threshold values for these bifurcations obtained from the model are slightly higher than those computed from DNS.

The results in a square box ($\eta = 1$) near the instability onset, computed from DNS as well as from the low-dimensional-model, are compared in Fig.~\ref{model_dns} for $Pr = 0.1$ and $Ta = 750$. The amplitudes of the temporal modulation of the Fourier modes $W_{101}$ [blue (black) curve] and $W_{011}$ [pink (gray) curve] just above the onset of convection ($r = 1.005$) are shown in the first column of Fig.~\ref{model_dns}. The modulation amplitudes (frequencies) of both the modes computed from the model [Fig.~\ref{model_dns} (b)] are within $8.5\%$ ($30\%$) of those obtained from DNS [Fig.~\ref{model_dns} (a)].  DNS gives larger period of amplitude modulation than that found in the model. As $Ra$ is raised for a fixed value of  $Ta$ in DNS, the quasiperiodic competition between two sets of rolls bifurcates into stationary square patterns at the secondary instability ($r =1.021$). 
The model also captures the stationary square patterns near the secondary instability. However, the secondary bifurcation  occurs at the higher value of $r$ in the model. The values of the two largest modes $W_{101}$ and  $W_{011}$ obtained from the model are compared with those obtained in DNS for $r= 1.025$. They are about $35\%$ smaller than those obtained from DNS [Fig.~\ref{model_dns} (c)]. The stationary square patterns ($W_{101} = W_{011}$) bifurcate into oscillating square patterns at $r = 1.306$, as $r$ is raised further. We observe qualitatively similar behavior in the model. The secondary and tertiary instabilities occur in the model at much higher values of $r$. The low-dimensional model qualitatively captures the essential features of the pattern dynamics near the onset of oscillatory convection in a slowly rotating Rayleigh-B\'{e}nard system with $\eta = 1$. The model becomes worse at higher rotation rates. The model and simulations suggest that the amplitude equations need to consider the distortions of fields even at the instability onset in low-Prandtl-number fluids in larger boxes and at higher rotation rates. 

\section{Conclusions}
We have investigated the effects of the Coriolis force on the convective patterns near onset of oscillatory convection in a Rayleigh-B\'{e}nard system ($Pr =0.1$) with stress-free top and bottom surfaces, and rotating uniformly about a vertical axis. We have considered several values of the horizontal aspect ratio $\eta$. For shorter boxes ($ 0.5 < \eta < 1 $ and $ 1 < \eta < 2$), the primary instability appears in the form of two-dimensional periodic standing waves for a wide range of the Taylor number $Ta$. The locations of up-flow and down-flow keep alternating periodically. The convective patterns at the primary instability in small containers of horizontal aspect ratios $\eta = 1$ and $2$ are interesting. The patterns at the primary instability in a square box ($\eta = 1$) depend on the value of $Ta$. The patterns may be due to  a competition between two sets of rolls having identical modulation amplitudes for smaller values of $Ta$, or due to a competition between two sets of rolls with unequal modulation amplitudes  for relatively higher values of $Ta$, or two-dimensional periodic standing waves for further higher values of $Ta$. The convective flow also shows stationary as well as oscillatory square patterns for $\eta =1$, if the Rayleigh number $Ra$ is raised keeping $Ta$ fixed. The patterns at the primary instability, in a rectangular box ($\eta = 2$), are due to quasiperiodic competition between two sets of rolls in mutually perpendicular directions for a wide range of $Ta$. The primary instability is always found to be chaotic in rectangular boxes with $4 \le \eta \le 10$. The patterns are due to a competition between two sets of rolls oriented at some angle in the horizontal plane.  These patterns have similarity with the pattern dynamics observed due to KL instability. However, the patterns near the onset of oscillatory convection appear due to a dynamical chaos rather than the noise as in the KL instability. The chaotic competition between a set of straight rolls and another set of oblique wavy rolls is initiated due to the excitation of the wavy modes in low-Prandtl-number fluids. Rhombic patterns are observed during the change in orientation of the rolls. The angle $\phi$ between the two sets of chaotically competing rolls is given by $\phi = \pi/2 - \arctan{(k_y/k_x)}$. 

{\bf Acknowledgements:} We benefited from fruitful discussions with Pinaki Pal, Priyanka Maity, Arnab Basak, and Deepesh Kumar. The presentation of this work has been greatly improved by useful comments from two anonymous referees.

\end{document}